# A Systematic Mapping Study of Crowd Knowledge Enhanced Software Engineering Research Using Stack Overflow

Minaoar Tanzil[a], Shaiful Chowdhury[b,*], Somayeh Modaberi[a], Gias Uddin[c] and Hadi Hemmati[c]

[a]University of Calgary, Calgary, Alberta, Canada
[b]University of Manitoba, Winnipeg, Manitoba, Canada
[c]York University, Toronto, Ontario, Canada

### ABSTRACT

Developers continuously interact in crowd-sourced community-based question-answer (Q&A) sites. Reportedly, ~30% of all software professionals visit the most popular Q&A site StackOverflow (SO) every day. Software engineering (SE) research studies are also increasingly using SO data. To find out the trend, implication, impact, and future research potential utilizing SO data, a systematic mapping study needs to be conducted. Following a rigorous reproducible mapping study approach, from 18 reputed SE journals and conferences, we collected 384 SO-based research articles and categorized them into 10 facets (i.e., themes). We found that SO contributes to 85% of SE research compared with popular Q&A sites such as Quora, and Reddit. We found that 18 SE domains directly benefited from SO data whereas *Recommender Systems*, and *API Design and Evolution* domains use SO data the most (15% and 16% of all SO-based research studies, respectively). *API Design and Evolution*, and *Machine Learning with/for SE* domains have consistent upward publication. *Deep Learning Bug Analysis* and *Code Cloning* research areas have the highest potential research impact recently. With the insights, recommendations, and facet-based categorized paper list from this mapping study, SE researchers can find out potential research areas according to their interest to utilize large-scale SO data.

## 1. Introduction

Software practitioners struggle to cope with the increasing complexity of modern software systems. Practitioners are demanded to keep pace with the continually changing technologies, which is often difficult and tiresome. Consider the case of learning a new programming language: even the expert practitioners often feel lost to learn and adapt to the new (and sometimes weird) conventions that are normal for a particular programming language [69]. To the practitioners' dismay, documentations about how to use these new technologies (e.g., a new language, a new framework, etc.) are often inadequate [21]. Consequently, different crowdsourcing platforms, such as Stack Overflow, have become popular among software practitioners, which help them to share knowledge with, and learn from, each other faster than ever before. According to the *similarweb*,[1] Stack Overflow (SO) is the most popular programming related question-answer (Q&A) site among software practitioners. As of writing, SO hosts 23 million questions, 34 million answers, and a staggering 19 million registered users[2]. Reportedly, ~30% of the software practitioners use SO every single day[3].

The popularity of SO is not limited to software practitioners anymore. The data produced through the enormous collaboration among the practitioners has made SO a productive platform for the SE research community as well. The pervasiveness of SO-based SE research is presented in Figure 1a. It shows the percent of research articles that used SO data compared to other popular Q&A sites, such as Reddit, and Quora (the detailed methodology and result are discussed in section 3).

By analyzing SO content—questions, answers, comments, code snippets, and community activities—researchers have produced a wide array of contributions in the field of software engineering. These contributions include understanding practitioners' expertise in certain topics [12, 34], building classifiers for developers' discussions [24], producing database of syntax errors and fixes [84], augmenting API documentation [76], understanding code quality [85, 48], and understanding gender issues in software engineering [47, 70], just to name a few. There is, however, no

---

[1], last accessed: 28-Aug-2023

[2]https://stackexchange.com/sites, last accessed: last accessed: 28-Nov-2023

[3]https://developerpitstop.com/how-often-do-software-engineers-use-stack-overflow/, last accessed: 14-Aug-2024

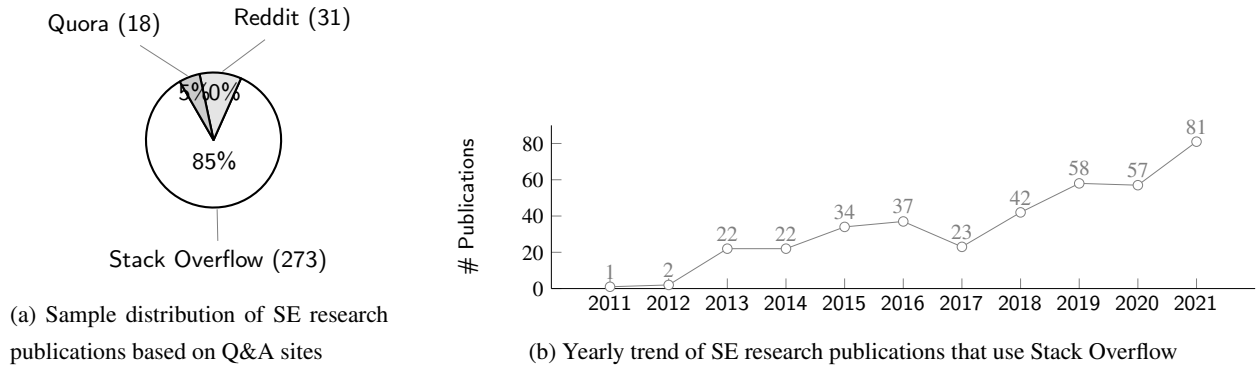

(a) Sample distribution of SE research publications based on Q&A sites

(b) Yearly trend of SE research publications that use Stack Overflow

**Figure 1:** Pervasiveness and trend of SO-based SE research compared to other popular community question-answer (Q&A) sites.

structured representation of how SO has been used by the research community: *What areas of the software engineering research have benefited the most (or the least) from SO? How the usage of SO has evolved in different software engineering research areas? And what areas of software engineering research are yet to take full advantage of SO?* Yes, there exists a preliminary study by Ahmad *et al.* [5] which attempted to answer similar questions related to SO research. That study, however, was limited to SO research papers that were published before June 2016. As shown in Figure 1b (details in section 3), most of the SO based research papers (69%) were published after 2016. Also, in that study, there is no investigation about the impact (e.g., citation) SO had in different areas of SE research. Therefore, there is a need to perform a new study to have better insights about the contributions that SO has made so far and can bring forth in the future to software engineering research.

In this paper, we perform a systematic mapping study of software engineering data-driven research that used SO as the primary data source. Our objective is to understand how SO has been used in SE research compared to other question-answer sites, and how the research community can take advantage of SO in future SE research. To achieve our objective, we have developed three research questions which are shown in Figure 2 and described as follows:

**RQ1. What is the trend and pervasiveness of SO-based SE research?** In a preliminary study, through multiple digital libraries and cross-reference searches, we found a total of 322 papers published in different reputed venues that studied three popular crowd-sourced community question-answer sites: SO, Quora, and Reddit. Among all these sites, a whopping 85% of studies are conducted using SO. There is also a strong trend in SO-based research since 2017.

Out of 384 papers in our mapping study, 266 papers (69%) were published since 2017, whereas only 118 papers were published before 2017. The last two full years of our study period (2020, and 2021) observed 138 published papers alone.

**RQ2. What are the facets (i.e., themes) that exist in SO-based SE research?** With a systematic approach, we have collected information about 384 software engineering papers that used SO data. We saved information about these papers according to 10 different facets (i.e., themes). For example, for every research paper, we stored the relevant SE domain (e.g., API Design and Evolution), the relevant phase of the software development life cycle (e.g., testing), the type of the research (validation, evaluation, or solution proposal), the research method, and so on. Such an indexed database of research papers according to their themes would be extremely helpful for a researcher who wants to know what SO-based studies are available on a certain topic of interest. For example, we have found that only 10% of the SO-based research is related to *software evolution*, whereas 50% of them is related to *software development*. Also, the most common research type is validation research (without industrial evaluation), the major research method is empirical research and most of the papers produce qualitative or descriptive models as their research contribution.

**RQ3. What SE domains are studied in SO-based SE research and what are their impacts?** Among the 10 different facets from RQ2, we found that the *SE domains* facet could be the most useful to the research community. Therefore, in this RQ we solely focus on the *SE domains facet*. We find that studies from 18 SE research domains have used SO data. Among them, *API Design and Evolution*,

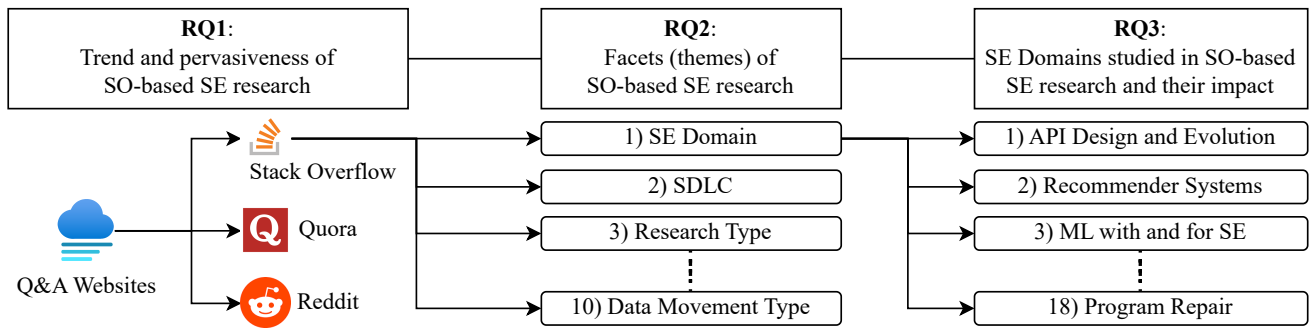

**Figure 2:** Research questions and outlines of the mapping study

*Recommender Systems*, and *ML with/for SE* have been extremely successful in using SO data. They constitute nearly half (181 out of 384 papers) of publications and also have higher average citations (more than 33 citations per paper). However, despite the high potential, some domains such as *Program Repair*, *Program Synthesis*, and *Requirement Engineering* did not leverage SO data significantly. In spite of the high impact factors (at least 32 citations per paper), the number of SO-based publications in all these three areas is considerably low (9 SO-based papers only in total).

To the best of our knowledge, this is the most comprehensive study that shows how SO has been used in software engineering research along with the gaps that exist therein. We believe that our study can guide the SE research community to better understand how to use SO in different SE domains, and which areas of software engineering research can be benefited the most in the future.

**Shared Data.** The data collection was an extremely laborious process which took the authors ∼200 hours for all the 384 papers—30 minutes per paper. To enable future research, we publicly share the summaries of these 384 papers in the replication package [1].

**Paper Organization.** Section 2 describes the process of paper collection, validation, and information extraction. Section 3, Section 4, and Section 5 answer RQ1, RQ2, and RQ3, respectively. Section 6 summarizes our findings and discusses the threats that can affect the validity of our findings. Section 7 discusses related works, and Section 8 concludes this paper.

## 2. Methodology

Petersen *et al.* [59] proposed a set of guidelines for performing systematic mapping studies in software engineering. Kitchenham *et al.* [41] provided similar guidelines for performing systematic literature reviews. Following Kitchenham's guideline, Brereton *et al.* [18] summarized the lessons for applying the systematic literature review process within the software engineering domain. The suggestions of these three studies have been followed in our study design which is outlined in Figure 3. At the beginning of the study, we prepared the protocol for the review process divided into preparatory and execution phases. In the preparatory phase, we defined the research questions and publication search and validation process including inclusion/exclusion criteria. Following this criteria and process definition, we then collected the final research papers. In the execution phase, we conducted a pilot review for defining the facets and their attributes. We then studied the collected research papers to collect data according to the defined facets and attributes. In the end, we performed data analysis to answer our research questions. We now describe each of the steps in Figure 3. To enhance readability, methods that are only relevant to a particular research question (RQ) are discussed with their corresponding RQ section only.

### 2.1. Paper Collection Process

Figure 4 summarizes our paper collection process, which is described as follows.

#### 2.1.1. Publication Search Strategy using Inclusion Criteria

Inspired by other similar studies [17, 19, 35, 7], we have selected five digital libraries to find and collect SO-based research papers: ACM Digital Library, IEEE Xplore Digital Library, Science Direct, Springer Link, and Google Scholar. We searched for the words "StackOverflow" and "Stack Overflow" within the titles and abstracts of the papers. Since the SO website was launched in 2008, we searched papers published within the timeline from January 2008 to February 2022. Unfortunately, all the libraries and web searches did not have uniform advanced search interfaces.

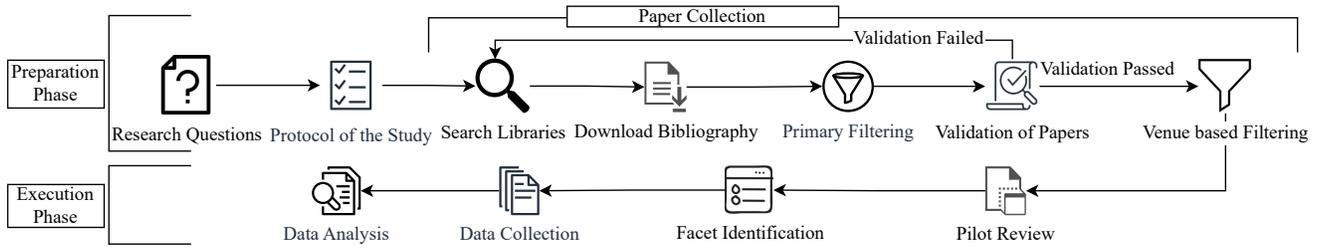

**Figure 3:** Phases and steps of the systematic mapping study

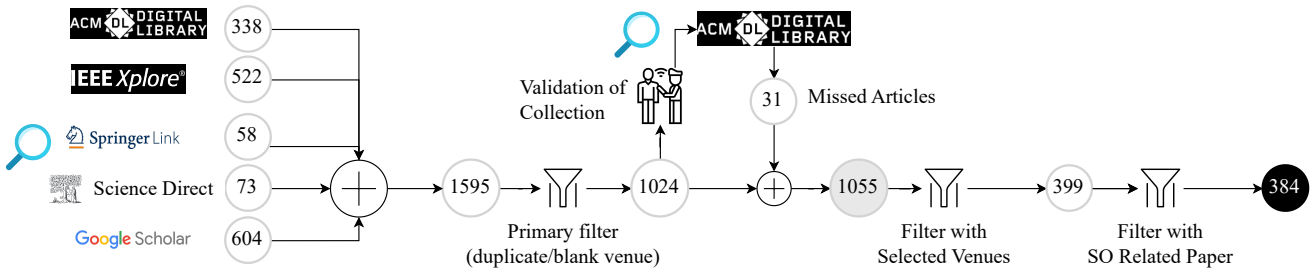

**Figure 4:** Detailed paper collection process for our mapping study. Initially, we collected 1,595 papers from the five libraries. However, after a systematic filter-out approach, we ended up with 384 research papers that we have studied further.

As a result, we had to change the search query according to the sources. The detailed search criteria are defined in Table 1. Since SpringerLink and Google Scholar provide an option to search only in titles, we could not search in abstracts. Moreover, only the ACM Digital Library and IEEE Explore provide options for month-level search granularity. For other libraries, the search period was, therefore, extended to 2023. Since we conducted this search at the end of February of 2022, putting the end year as 2023 ensured that all publications till February of 2022 were reported, hence making the same search period for all libraries. The digital libraries (SpringerLink and Google Scholar) did not explicitly document whether the search results included or excluded the end-year criteria. Hence, to be on the safe side, instead of putting the end-year 2022, we used 2023 as the end-year criteria. The search process returned a total of **1,595** potential research papers. Google Scholar provided the most (604) and SpringerLink returned the least (58) numbers of papers in their such results.

### 2.1.2. Downloading Bibliography from Search Results

We could not find a generic approach to download and save publication information about the 1,595 research papers found in the library search. Each of the libraries provides different methods to export the bibliographic list. For example, SpringerLink provides a comma-separated values (CSV) file with the list of papers whereas ACM Digital Library, IEEE

**Figure 5:** Google Scholar results showing publication citations without linking to any online URL of the publication

Explore, and ScienceDirect provide the paper citations in `bib` format. For Google Scholar, we had to manually select and add each paper to a favorite paper list and then downloaded the bibliographic reference of all papers together in a CSV file. For the citations from ACM Digital Library, IEEE Explore, and ScienceDirect, we converted the bib format files to CSV using the BibTex Converter[4]. In the end, we produced some CSV files from all five sources. We then merged all the files into a single CSV file.

### 2.1.3. Primary Exclusion Criteria

We excluded any non-English language papers that we found from Google Scholar. Moreover, in Google Scholar, a large number of the search results were actually citation links to the papers, not references to actual papers. We, therefore,

---

[4] http://bibtex.com/

**Table 1**

Search criteria across digital publication libraries

| Library | Period | Search Area | Advanced Search Query |
|---|---|---|---|
| ACM Digital Library | Jan08-Feb22 | Title, Abstract | Title:(stackoverflow or "stack overflow") OR Abstract:(stackoverflow or "stack overflow") |
| IEEE Xplore | Jan08-Feb22 | Title, Abstract | ("Abstract":"stackoverflow" OR "Abstract":"stack overflow") OR ("Document Title":"stackoverflow" OR "Document Title":"stack overflow") |
| ScienceDirect | 2008-2023 | Title, Abstract | Title, abstract keywords:("stackoverflow" OR "stack Overflow"), Year: 2008-2023 |
| SpringerLink | 2008-2023 | Title | "stackoverflow", "stack overflow" |
| Google Scholar | 2008-2023 | Title | allintitle: stackoverflow OR "stack overflow" |

excluded these citation links. Examples of such citation-only links are shown in Figure 5. For example, the paper "Mining Duplicate Questions in Stack Overflow" [4] is listed in the results with [CITATION] tag in front and the result does not provide any URL to access the publication. Moreover, this paper is also listed another time among the 604 publications, but with proper links. Hence, when we discarded such [CITATION] results from Google Scholar, we did not lose any publication. Unlike the other four libraries, Google Scholar contained poster papers, which we also excluded from our study. At this stage, we had 499 research papers from the Google Scholar library and a total of 1,490 research papers from the five libraries. However, not surprisingly, many of these papers were duplicates —e.g., the same paper was found both in the ACM and IEEE libraries. After removing those duplicates, the number of papers was reduced to **1,024**.

### 2.1.4. Validation of Paper Collection

We should now ask, *does our list of 1,024 papers contain all the high-quality SO-based SE studies?* We can never be 100% certain, but with a validation process, we can understand how robust was our paper collection process, and if there is any way to improve it even further. According to our paper collection protocol, the paper list was collected and prepared by the first author. The third and the fourth author conducted a completely independent search and prepared a validation list of 20 papers and a testing list of 10 papers. Besides searching, they also applied their experiences to include research papers that they considered important. The 20 papers of the validation list were then shared with the first author.

The first author found 18 of the 20 validation papers in the master list of 1,024 papers. They then investigated and found the root cause for the two missing papers: in the ACM digital library, there were some research papers listed only as 'article', not as 'research article'. For the ACM digital library, our search process was to look for '*research articles*' only—to avoid posters, proposals, etc. For unknown reasons, however, these two papers were listed as '*article*' only (although they were full research papers). So the first author changed the search criteria in ACM Digital Library by removing the filter of '*Research Article*', which resulted in 72 more papers that are not in our master list. After manual reviewing, however, the first author found that only 31 of them were actual research papers; the rest of the 41 papers were posters and proposals. They, therefore, added these 31 papers to the master list, resulting in **1,055 papers**.

At this stage, the first author was provided with the final test list of 10 papers. This time the master paper list contained all 10 papers (100% hit). We are, therefore, extremely confident that our master list of 1,055 papers contains all the relevant SO-based SE research papers.

### 2.1.5. Exclusion Criteria based on Publication Venue

Although the validation process ensured that we have a comprehensive list of publications related to SO, unfortunately, many of the collected research papers do not fit with software engineering research. Also, the review quality of some of the venues is unknown to us. On the other hand, we wanted to study high-quality software engineering research papers only. To alleviate these two problems, we developed a list of venues (i.e., journals and conferences) that are connected to software engineering research only and have a high reputation within the research community. This list was developed in two steps. First, we collected the top 20 venues according to the Google Scholar within

**Table 2**

List of selected publication venues. *SRC* column shows the criteria why a venue was selected: GS means the venue was found in the top venue list provided by Google Scholar, and R means the venue was provided by the second and the fourth author. *C#* column denotes the number of papers from a venue. (C) or (J) in the publication name denotes the venue types Conference and Journal respectively.

| Publication | SRC | C# |
|---|---|---|
| 1. (C) ACM/IEEE International Conference on Software Engineering (ICSE) | R+GS | 69 |
| 2. (J) Journal of Systems and Software (JSS) | R+GS | 18 |
| 3. (C) IEEE Transactions on Software Engineering (TSE) | R+GS | 28 |
| 4. (J) Information and Software Technology (IST) | R+GS | 12 |
| 5. (C) ACM SIGSOFT International Symposium on Foundations of Software Engineering (FSE) | R+GS | 28 |
| 6. (J) Empirical Software Engineering (EMSE) | R+GS | 16 |
| 7. (J) IEEE Software | R+GS | 2 |
| 8. (C) Mining Software Repositories (MSR) | R+GS | 82 |
| 9. (C) IEEE/ACM International Conference on Automated Software Engineering (ASE) | R+GS | 24 |
| 10. (C) International Conference on Software Analysis, Evolution, and Reengineering (SANER) | R+GS | 27 |
| 11. (C) International Symposium on Software Testing and Analysis (ISSTA) | R+GS | 2 |
| 12. (C) IEEE International Conference on Software Maintenance and Evolution (ICSME) | R+GS | 39 |
| 13. (C) IEEE/ACM International Conference on Program Comprehension (ICPC) | R | 10 |
| 14. (C) Empirical Software Engineering and Measurement (ESEM) | R | 12 |
| 15. (C) IEEE International Conference on Software Testing, Verification and Validation (ICST) | R | 2 |
| 16. (J) International Working Conference on Source Code Analysis and Manipulation (SCAM) | R | 6 |
| 17. (J) ACM Transactions on Software Engineering and Methodology (TOSEM) | R | 8 |
| 18. (C) Advances in Social Networks Analysis and Mining (ASONAM) | R | 14 |
| 19. (C) International Conference on Fundamental Approaches to Software Engineering (FASE) | R | 0 |
| 20. (J) Journal of Software: Evolution and Process | R | 0 |
| 21. (J) Software Testing, Verification and Reliability (STVR) | R | 0 |
| 22. (C) Tools & Algorithms for the Construction & Analysis of Systems (TACAS) | GS | 0 |
| 23. (C) ACM SIGPLAN Symposium on Principles and Practice of Parallel Programming (PPOPP) | GS | 0 |
| 24. (C) ACM SIGPLAN-SIGACT Symposium on Principles of Programming Languages (POPL) | GS | 0 |
| 25. (J) Proceedings of the ACM on Programming Languages (PACMPL) | GS | 0 |
| 26. (J) Software and Systems Modeling (SoSyM) | GS | 0 |

the category "Software Systems".[5] We later found this list to be inadequate. For example, to our surprise, it did not include the "ACM Transactions on Software Engineering and Methodology", which is a highly prestigious journal to the SE research community. To enlarge our selected venue list, the second and the fourth author had a one-hour meeting to make a list of venues that they considered as reputed venues for publishing software engineering research papers. These two authors have Ph.D. in software engineering and have published a significant number of research papers using SO-data. In the end, after merging the two lists, we have produced a final list of 26 venues, as presented in Table 2.

After selecting papers published only in these 26 SE venues, the number of papers became **399** from 1,055. As presented in Table 2, only 18 of the 26 venues have published these 399 papers.

### 2.1.6. Final Screening of Papers

Before starting the data collection, we went through the title and abstract of each article and excluded a few papers which were duplicates or which did not use SO data. After excluding such non-relevant papers, the final count of the articles of the study came down to **384**.

We identified that a few papers have been duplicated that we could not identify earlier. When we merged all paper lists from different digital libraries, we identified the duplicate papers by their titles. However, during our final study phase, we realized that few libraries had different titles listed under different libraries. For example, IEEE Explore listed a paper with the title "Which Non-functional Requirements Do Developers Focus On? An Empirical Study on Stack Overflow Using Topic Analysis" [88], whereas ACM Digital Library listed the same paper with a truncated title "An Empirical Study on Stack Overflow Using Topic Analysis". Sometimes different digital libraries used different punctuation in the title of the papers. For example, IEEE Explore used a hyphen in the title "One-Day Flies on StackOverflow - Why the Vast Majority of StackOverflow Users Only Posts Once" whereas ICM digital library used a colon in their title "One-Day Flies on StackOverflow: Why the Vast Majority of StackOverflow Users Only Posts Once" [71].

Few papers mentioned Stack Overflow in their abstract but did not actually refer to the website Stack Overflow, rather discussed the overflow of stack data structure. For example, "Expositor: Scriptable Time-Travel Debugging with First-Class Traces" [60].

Some papers mentioned Stack Overflow in their abstract just as a reference but did not actually conduct the study around stack overflow. For example, the article "SOA4DM: Applying an SOA Paradigm to Coordination in Humanitarian Disaster Response" studied the applicability of service-oriented architecture to tackle humanitarian disaster response and just mentioned SO and GitHub as popular collaboration sites without actually using these sites in their research [43]. Another example of such an article that only referred to SO as an example in their abstract is "BugsInPy: A Database of Existing Bugs in Python Programs to Enable Controlled Testing and Debugging Studies" [80].

After this filtering process, it is still possible that we missed some of the useful SO-related papers but without such filtering, it would be very difficult to go through all the 1,055 papers. Moreover, we can investigate the research questions on these 384 papers since each of these papers is published in reputable peer-reviewed conferences and journals. All the results and analyses in subsequent sections are prepared based on our final **384** articles.

## 2.2. Pilot Review

We needed to read and collect data from a staggering 384 research papers. Repeating this phase more than one time would be expensive. Therefore, it was crucial for us to establish a framework before starting the paper-reading phase. To establish the framework, we conducted a pilot study as suggested by Kitchenham *et al.* [41] and Brereton

*et al.* [18]. As we have 384 papers, we wanted to conduct the pilot study on around 10% of those papers. So, we chose a target paper count of around 40. We plot a frequency table with the number of papers per year and found that 49.5% of papers are published in or after 2019. So we decided that in the pilot review, around 50% papers should come from 2019 onward. We also sorted the venue list with the highest number of papers and found that the top 10 venues published 345 papers (90% of all 384 papers). So we selected papers from the top 10 venues only. Then for each venue, we decided to select papers by the weighted average of their number of publications. For example, out of the 345 papers from 10 venues, International Conference on Mining Software Repositories (MSR) published 80 papers. So in our 40-paper list, we selected 10 papers from MSR— 82/(345*40) ≈10. Then out of these 10 MSR papers, we selected 5 papers from 2019 onward, and 5 papers from before 2019. In this way, we finally selected 44 papers from 10 venues. An extra 4 articles came because of the ceiling of the fraction. The paper distribution for the pilot review is shown in Table 3.

## 2.3. The Attribute Framework

Because of the large number of research papers with diverse topics, it is very important to set up a data collection framework at the very beginning of our mapping study. We found the attribute framework by Cornelissen *et al.* [25] to be useful for collecting and analyzing multi-faceted data that we have. In the attribute framework, Cornelissen *et al.* identified a few facets (such as *activity*, *method*, and *evaluation* performed in a research paper), and under each facet, they collected specific attributes (for example, case study, industrial comparison, etc. were the attributes of the facet *evaluation*) of each particular research paper.

Encouraged by the study of Cornelissen *et al.* [25], we have built an attribute framework to systematically characterize each of our 384 research papers. In this framework, each of the 384 research papers will be decomposed into a set of facets (i.e., themes). Within each facet, there exists a set of generic attributes. Every research paper will be associated with one of the attributes within a facet. For example, the research paper "Stack Overflow: A code laundering platform?" by An *et al.* [8] fits with the "Ethics in Software Engineering" attribute within the "SE Domain" facet.

To develop the complete set of facets and their finite set of attributes, the first author studied all 44 papers that were selected for the pilot study. To improve the attribute selection

**Table 3**

Pilot review papers: year and venue wise distribution. P column shows the number of papers taken in the pilot, 2019− denotes papers prior to 2019, and 2019+ denotes the number of papers on or after 2019.

| # | Publication Venue | All | P | 2019− | 2019+ |
|---|---|---|---|---|---|
| 1 | Mining Software Repositories (MSR) | 82 | 10 | 5 | 5 |
| 2 | ACM/IEEE International Conference on Software Engineering (ICSE) | 69 | 8 | 4 | 4 |
| 3 | IEEE International Conference on Software Maintenance and Evolution (ICSME) | 39 | 5 | 3 | 2 |
| 4 | IEEE Transactions on Software Engineering (TSE) | 28 | 4 | 2 | 2 |
| 5 | Software Analysis, Evolution, and Reengineering (SANER) | 28 | 4 | 2 | 2 |
| 6 | IEEE/ACM International Conference on Automated Software Engineering (ASE) | 27 | 3 | 2 | 1 |
| 7 | Symposium on Foundations of Software Engineering (FSE) | 24 | 3 | 2 | 1 |
| 8 | Journal of Systems and Software (JSS) | 18 | 3 | 2 | 1 |
| 9 | Empirical Software Engineering (EMSE) | 16 | 2 | 1 | 1 |
| 10 | Advances in Social Networks Analysis and Mining (ASONAM) | 14 | 2 | 1 | 1 |
| | Total Papers | 345 | 44 | 24 | 20 |

process, the first author consulted relevant guidelines on systematic literature review [41], mapping studies [59, 81], and research recommendations on software engineering [83]. After the analysis of the pilot study, we found the facets and their attributes. For example, SE Domain was a prominent facet with 18 attributes such as API Design and Evolution, Recommender Systems, and Machine Learning with and for SE. Another facet was to identify which of the Software Development Lifecycle Stages was focused on in a paper. This facet had five attributes - *specification*, *development*, *validation*, *evolution*, and *generic (no specific)* stage. The definition of each facet and their corresponding attributes is elaborated in section 4.

An open coding [50] approach was followed to label the attributes found against each publication. Open coding is a qualitative analytical process to label concepts found in observational data. The labels attached to each concept are termed as a 'code' and the process of such labeling is called open coding. The first author then discussed (and modified when needed) all the facets and attributes with the second and the fourth authors to finalize the attribute framework. The first and the third author were responsible to read and extract information from all the 384 papers. Therefore, the first author explained the whole framework to the third author to make it clear how to read, extract information and label the information from a research paper.

To evaluate if the first and the third authors are on the same page while labeling a research paper, we calculated Cohen's Kappa ($k$) coefficient for measuring the inter-rater reliability on five selected facets (all the facets are described later). Table 4 presents the results. While labeling the attributes, we found substantial agreement for the *Research Objective* and *Beneficiary* facets, since they are comparatively straightforward. For the *SO Data* and *SO Content Usage*, the agreement was moderate and the agreement for labeling the *Data Movement Type* facet was fair. The first and the third authors then discussed the mismatches and came to a consensus on how they should label in the future to unify the coding, before starting to read and label all of the 384 research papers.

## 2.4. Data Collection and Analysis

With the attribute framework prepared, the first and third authors divided the research papers among themselves to study. Then we studied the papers and stored the attributes of each paper based on the 10 facets. In addition to the specific facets, we also kept short notes on each paper for future reference. The final data collection sheet contained the facets, notes, and title, abstract, author, venue, and citation count (according to Google Scholar) for 384 SO-based SE papers.

In section 4 and 5, we present the result and analysis of the collected data based on the attribute framework.

## 3. (RQ1) What is the Trend and Pervasiveness of SO-based SE Research?

We collected 384 peer-reviewed articles published in reputed SE conferences and journals. However, there could be other crowd-sourced community question-answer platforms that might have been studied in the SE research community as well. Hence, we first explored this pervasiveness of the

**Table 4**

Inter-rater reliability for labeling attributes in four selected facets. The meaning of these facets is described later in Section 4

|  | SO Content Usage | SO Data | Beneficiary | Research Objective | Data Movement |
|---|---|---|---|---|---|
| % of agreement | 85% | 80% | 90% | 85% | 70% |
| Cohen's Kappa | 0.48 | 0.47 | 0.61 | 0.66 | 0.28 |
| Inter-rater agreement | Moderate | Moderate | Substantial | Substantial | Fair |

SO-based studies with a question to see if SO really stands out compared to other sites regarding SE research.

### 3.1. Pervasiveness of SO-based SE research

#### 3.1.1. Comparison of SO with Popular Question-Answer Sites in SE Research.

To see how SE researchers study crowd-based question-answer sites, we first chose three popular sites: Quora[6], Reddit[7], and Stack Overflow (SO) [8]. To make a comparison among these three sites, it would be extremely time-consuming if we follow the paper collection process for Quora, and Reddit similar to the process for SO. As we are not summarizing research papers that used Quora, and Reddit, rather we are trying to have a quick estimated comparison among these three sites, the accuracy of the paper collection process does not have to be very high.

To search research articles based on these sites, we used a tool 'Publish or Perish'[9], developed by Harzing [36]. The Publish and Perish tool provides a desktop-based application. It offers a search interface for popular search and research indexing sites such as OpenAlex[10], CrossRef[11], Scopus[12], Google Scholar[13]. Before running thorough searches for community site-based research articles, we needed to identify which publication indexing service provides the most result through the Publish or Perish interface. Hence, we provided a fixed search word 'stackoverflow' for each indexing engine. We found that CrossRef engine returned 73 papers, Google Search returned 197 papers, OpenAlex provided 667 papers, and Scopus provided 75 papers. Since OpenAlex returned the maximum number of articles, we selected OpenAlex as the indexing engine for subsequent search results.

We searched OpenAlex with the following search strings: *reddit*, *quora*, *stack overflow*, *stackoverflow*, *community question answer*, *crowdsource*, and *software crowdsource*. We downloaded each search result as a comma-separated file (CSV) that contains the paper title, publication venue, publication year, citation count, and abstract. We merged all the data files and kept the unique papers (based on the title). If any paper is published in the reputed SE venues, we considered the paper as a SE research study. We used the same SE venue list that we used to collect SO-based articles and we also added two more venues (arXiv: Software, arXiv: Comput) so that we can cover as many articles as possible (not only peer-reviewed published articles). Since our goal is to identify which community question-answer site is used most in the research paper, we searched the abstract of each article to find the name of the community site. If the abstract contained names of any of the community sites (Reddit, Quora, SO), we considered the study to be based on that community site. The result of our analysis is shown in Table 5.

With all the search terms, we found a total of 8,749 unique papers, and out of those papers, 637 articles were published in SE venues (~7% papers). We then manually studied these 637 papers and found that only 322 papers actually used these three sites, and the others were false positives. Among these 322 papers, 273 studied SO, 31 papers studied Reddit, and 18 papers studied Quora. Figure 1a compares the distribution and confirms the prevalence of SO-based studies in SE research among the selected question-answer sites. We also found that SO-based papers have the most average citation (22) followed by Reddit (19).

#### 3.1.2. Prevalence of SO-based Research in Reputed SE Venues.

We also analyzed the presence of SO-based research studies in reputed SE venues based on the selected 384 papers of the mapping study. Figure 6 shows the number of publications and average citations separately for conferences and journals. In addition to the higher number of citations,

---

[6] https://www.quora.com
[7] https://www.reddit.com
[8] https://stackoverflow.com
[9] https://harzing.com/resources/publish-or-perish
[10] https://openalex.org
[11] https://www.crossref.org
[12] https://www.scopus.com
[13] https://scholar.google.com

**Table 5**

SE research papers based on the three selected community question-answer sites

| Search Term | Total papers | SE papers | SO Papers | Quora Papers | Reddit Papers |
|---|---|---|---|---|---|
| *stackoverflow* | 1,498 | 478 | 251 | 2 | 0 |
| *quora* | 1,122 | 50 | 18 | 17 | 1 |
| *reddit* | 2,390 | 81 | 3 | 0 | 30 |
| *crowdsource* | 3,931 | 60 | 16 | 0 | 0 |
| *codeproject* | 38 | 3 | 3 | 0 | 0 |
| Total | 8,979 | 672 | 291 | 19 | 31 |
| Unique Papers | 8,749 | 637 | 273 | 18 | 31 |
| Average Citation | | 35 | 22 | 6 | 19 |

SO-based research papers are published more frequently in conferences than in journals. This seems to align with the previous findings [30] that in computer science, conferences often have equal or more impact than journals.

Table 6 is the more elaborate depiction of the 18 different venues that published SO-based SE research. Unsurprisingly, the *International Conference on Mining Software Repositories (MSR)* has published the highest number of SO-based research papers. SO-based research fits the best with the MSR objectives, so much so that research based on "question-and-answer sites" is indicatively mentioned in the MSR call for paper section.[14] MSR also has a dedicated mining challenge track that multiple times invited research papers that solved SE research problems utilizing SO-data.[15][16]

The staggering 18 different venues containing SO-based research imply that SO has attracted researchers from different SE communities. While the list of venues includes conferences and journals that accept research from all the areas of software engineering (e.g, ICSE, ESEC/FSE, TSE, and TOSEM), SO has also been used by communities that are more focused on specific research areas—social network analysis (e.g., ASONAM), Testing (e.g, ISSTA and ICST), and program comprehension (e.g., ICPC). These observations encouraged us to have more insights into the characteristics of SO-based SE research (RQ2 and RQ3) so that we can better understand the past to guide future SO-based research.

---

[14] `https://conf.researchr.org/track/msr-2023/msr-2023-technical-papers` : last accessed Jan 01, 2023

[15] `https://2015.msrconf.org/challenge.php`: last accessed Jan 01, 2023

[16] `https://2013.msrconf.org/challenge.php`: last accessed Jan 01, 2023

## 3.2. Trend of SO-based SE research

Figure 7 shows the trend of SO-based SE research publications based on the 384 articles we selected. The first research paper was published in 2011, just two years after SO was founded in July 2008. Since then, with few fluctuations, the number of publications increased steadily. The highest number of 81 papers were published in the most recent full year of 2021. The number of publications in 2022 is low because our paper collection stopped in February 2022. We can see that 69% of the research papers was published after 2016, making previous related studies (e.g.,[5, 49]) partially obsolete.

---

**RQ1 Summary:** Stack Overflow has become instrumental in enhancing data-driven software engineering research, and its pervasiveness is ever-increasing. Compared to other popular question-answer sites (Quora, Reddit), SO is the most prevalent in SE research studies: 85% of community question-answer site-related SE studies are related to SO only. Apparently, SO is being utilized by diverse SE research communities. This motivated us to push the study even further to have more insights into the impact that SO had (and can have in the future) in different SE areas, which we investigate in RQ2 and RQ3.

---

## 4. (RQ2) What are the Facets (i.e., themes) that Exist in SO-based SE Research?

In the previous section, we have established the prevalence of SO in SE research. Our objective now is to classify SO-based studies into different facets and attributes, which may work as an indexed database for the research community. After completing the labeling phase described in Section 2.3, we were able to label each research paper

**Table 6**

Venue-wise publication count and average citation. Black=publication, Grey=Average citation. The number of citations of each paper was collected from Google Scholar [67]. ISSTA has the most average number of citations, but it contains only two research papers. With 16 papers and 67 average citations per paper, EMSE stands out among all the venues.

| Publication Venue | Paper Count \| Average Citation |
|---|---|
| 1. Int. Conf. on Mining Software Repositories (MSR) | 80 46 |
| 2. Int. Conf. on Software Engineering (ICSE) | 64 47 |
| 3. Int. Conf. on Software Maintenance and Evolution (ICSME) | 38 37 |
| 4. Int. Symposium on Foundations of Software Engineering (FSE) | 27 29 |
| 5. Transactions on Software Engineering (TSE)(J) | 27 15 |
| 6. Int. Conf. on Software Analysis, Evolution, and Reengineering (SANER) | 26 39 |
| 7. Int. Conf. on Automated Software Engineering (ASE) | 23 26 |
| 8. Journal of Systems and Software (JSS) | 17 8 |
| 9. Empirical Software Engineering (EMSE) | 16 67 |
| 10. Int. Conf. on Advances in Social Networks Analysis & Mining (ASONAM) | 14 31 |
| 11. Int. Symp. on Empirical Software Engineering & Measurement (ESEM) | 12 17 |
| 12. Information and Software Technology (IST) | 12 36 |
| 13. Int. Conf. on Program Comprehension (ICPC) | 10 37 |
| 14. Transactions on Software Engineering and Methodology (TOSEM) | 7 10 |
| 15. Int. Working Conf. on Source Code Analysis & Manipulation (SCAM) | 6 11 |
| 16. IEEE Software | 2 51 |
| 17. Int. Symposium on Software Testing and Analysis (ISSTA) | 2 77 |
| 18. Int. Conf. on Software Testing, Verification and Validation (ICSTW) | 1 |

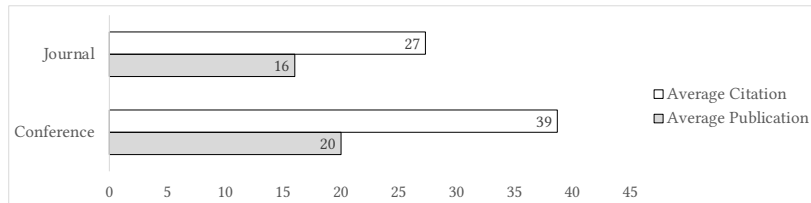

**Figure 6:** Research and citation distribution by Publication Venues Types

with an attribute under each of the 10 facets. Table 7 shows the summary of each facet with attribute examples. The minimum and the maximum number of attributes are three and eighteen for the Research Types and SE Domains facets, respectively. We publicly share this attribute framework so that future SO-based research can immediately find all the research papers related to a specific facet or attribute [1]. For example, as we discuss later, the *code analysis* attribute under the *research objective* facets include all the research papers that leveraged SO code snippets to understand and

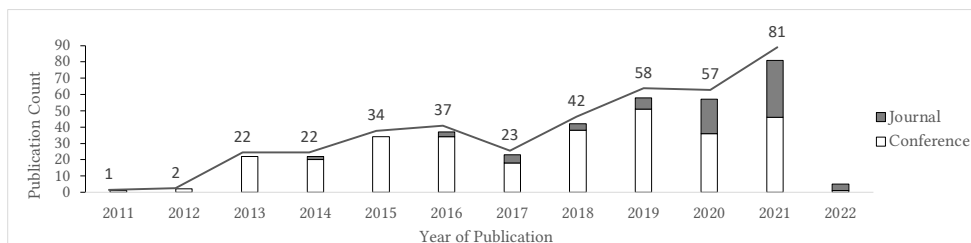

**Figure 7:** Research distribution by years

**Table 7**

All the facets, their description, and top attributes. Column *A#* denotes the number of attributes in the facet.

| # | Facet | Description | A# | Example attributes |
|---|-------|-------------|-----|-------------------|
| 1 | SE Domain | Which SE domain is targeted in the research? | 18 | API Design and Evolution, Reco. Systems |
| 2 | SDLC | Is any specific SDLC phase targeted? | 5 | Development, Generic |
| 3 | Research Type | What type of research has been conducted? | 3 | Validation Research, Evaluation Research |
| 4 | Contribution Type | What is the contribution of the study? | 3 | Prototype, Procedure |
| 5 | Research Method | What research method(s) applied in the study? | 6 | Empirical, Machine Learning |
| 6 | Research Objective | What is the primary objective of the research? | 7 | Tool Development, Topic Analysis |
| 7 | SO Content Usage | How the content of SO has been used in the study? | 3 | Content Analysis, Content Extraction |
| 8 | Beneficiary | Who are the primary beneficiary of the study? | 3 | Researchers, Developers |
| 9 | SO Data Field | Which particular SO data has been used in the study? | 12 | Post, Code Snippet |
| 10 | Data Movement Type | How the data from/to SO system is moved and used? | 3 | Internal (SO) to Internal, Int. (SO) to Ext. |

solve diverse software engineering research problems. Next, we summarize each facet to help the research community understand more about all the facets and their attributes.

**(1) SE Domain.** Each of our 384 research papers belongs to a specific domain (an attribute). These domains (i.e., attributes for the facet SE Domain) were defined during the labeling phase, described in Section 2.3. In that phase, we first made a unique set of domains by merging all the domains from the *2022 call for paper* sections from five selected software engineering conferences: ICSE[17], ESEC/FSE[18], ASE[19], ICSME[20], and MSR[21]. These five venues were selected after a discussion among the first, second, and fourth authors. The idea was to label a paper to a domain that best describes its methods or contribution. For example, the domain of the paper "Mining Architecture Tactics and Quality Attributes knowledge in Stack Overflow" by Bi *et al.* [16] was labeled as "Software Architecture and Design". In contrast, the domain of the paper "Usage and attribution of

Stack Overflow code snippets in GitHub projects" by Baltes *et al.* [10] was labeled as "Ethics in Software Engineering". In the end, all 384 papers were labeled using 18 domains (attributes). Although a paper can match multiple attributes, we selected only one attribute that describes the paper the most. We also noticed that these 18 domains can be merged into higher-level research categories, and can be divided into more lower-level research sub-domains. We describe these areas and sub-domains in section 5 when we answer RQ3 in depth.

**(2) Software Development Lifecycle (SDLC).** We identified what are the distribution of the research works among different stages of SDLC. Since there are various software activities in software lifecycle models (waterfall model, incremental development model), we followed basic process activities of *specification*, *development*, *validation*, and *evolution*—as was defined by Sommerville [72]. If a research paper is specifically focused on any of these activities, we assigned that activity to that paper as its attribute. For example, we labeled the paper "Should We Move to Stack Overflow? Measuring the Utility of Social Media for Developer Support" [73] with the *evolution* attribute, because developer support is part of *evolution* which deals with software maintenance. There are, however, some research papers that do not fit with any of these four activities. We labeled those papers with the *generic* attribute. For example, the paper "Mining stackoverflow to filter out off-topic irc discussion" [24] was labeled as *generic*, because it proposes a research-domain agnostic two-dimensional fixed classi-fier to filter out off-topic discussions from developers' forums, which can be relevant to any of the four lifecycle activities. Figure 8 shows the distribution of research papers in these five categories. 50% of the publications are focused on the development stage followed by 33% generic articles which do not cover any particular lifecycle stage. Software specification has the least number of publications (2%).

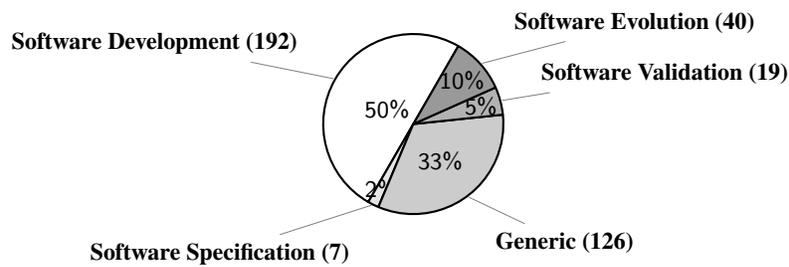

**Figure 8:** Distribution of publications focused on different life cycle stages of software development

**(3) Research Types.** As we would like to provide an overview of SE studies using SO, it is important to classify the studies following a standard classification. In guide-lines for mapping studies, Petersen *et al.* [59] suggested a research-domain agnostic two-dimensional fixed classi-fication scheme. One important dimension of the classi-fication scheme is the Research Type of the articles (the other dimension is the Contribution Type which is discussed later). Petersen *et al.* [59] referred to Wieringa *et al.* [81] for the definition and classification of the Research Types. Following the classification, we categorized all the papers into three research types: *solution proposal*, *validation*, and *evaluation*.

According to the definition by Wieringa *et al.* [81], and adaptation of Petersen *et al.* [59], an article falls into the solution proposal category when a solution design, to improve the current situation, is proposed without any im-plementation and empirical evidence. For example, Bayati [15] proposed a solution design to identify security experts by analyzing their SO posts. They, however, did not conduct any empirical study on SO to identify the experts.

An article falls into the category of validation research when it empirically studies the problem area or provides a laboratory implementation of a proposed solution without evaluating it in the industry. Saha *et al.* [66] analyzed unan-swered SO question answers to find out the reason for the questions that remained unanswered. Through this empirical study, they also identified primary reasons for this problem but they did not evaluate their findings by surveying or interviewing any active or renowned SO user. Hence, this study falls under the validation category.

Finally, evaluation research is the investigation or imple-mentation of a solution in the actual industrial practice [81]. For example, Huang *et al.* [37] surveyed developers on how they use SO for API-related knowledge gathering. Then they designed and implemented a solution to help the developers and finally conducted a user study on their implemented approach. Because of this final user study in the industry, this article falls under the category of evaluation research.

As shown in Figure 9a, most articles are validation research (81%) followed by evaluation research (17%). There are still 2% of articles providing solution proposals in short paper formats.

**(4) Contribution Types.**

Contribution Types is the other dimension of the domain-agnostic fixed classification scheme suggested by Petersen *et al.* [59]. To analyze the contribution type of a research paper, we followed the guidelines provided by Shaw [68]. We labeled the contribution type of a research paper with one of the three broad categories: *qualitative or descriptive model*, *procedure or technique*, and *tool or prototype*. According to the definition of Shaw, we considered a research paper within the *qualitative or descriptive model* category, if it provides a qualitative or quantitative analysis result or empirical investigation report without proposing any specific solution or technique. For example, Ragkhitwetsagul *et al.* [62] identified SO code snippets that are either outdated from their original project or which violate the software license. This study does not provide any particular procedure or tool for detecting such toxic code, rather provides the output

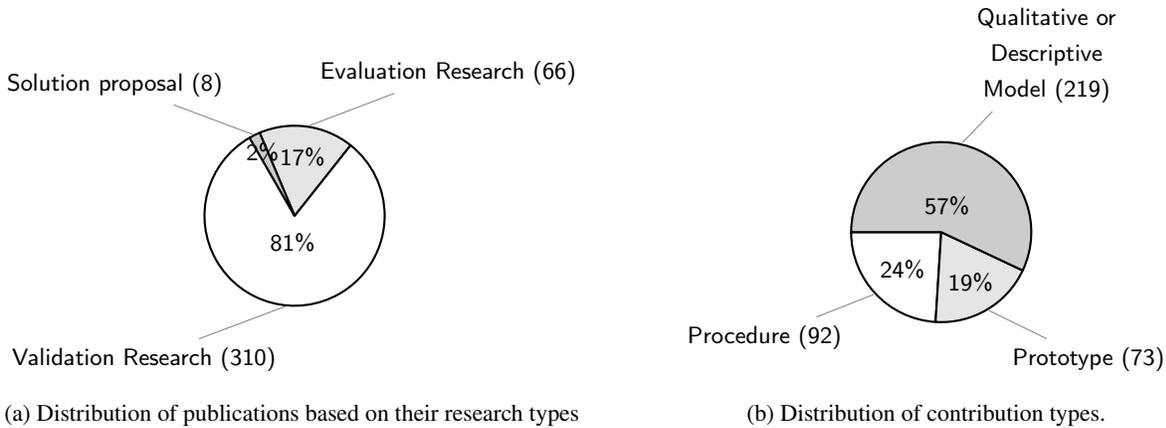

(a) Distribution of publications based on their research types

(b) Distribution of contribution types.

**Figure 9:** Distribution of facets research types and contribution type

of their analysis. We labeled a research paper with the *procedure or technique* category if a new or better way to do some task, such as design, or implementation is proposed. Sometimes they can also include operational techniques for implementation, representation, or management of an existing problem area. For example, Ponzanelli *et al.* [61] presented an automated post-improvement method by considering the post content and poster profile in SO. Since this provides a technique to improve the problem of low-quality posts, its contribution was labeled as procedure or technique. Finally, we considered the outcome as a *prototype* if a specific new tool is developed or enhanced, even if it was not evaluated in practice. For example, Vassallo [77] developed a plugin for Eclipse, an integrated development environment (IDE), named CODES to create the Javadoc descriptions from SO comments regarding particular libraries. Figure 9b shows the distribution of papers in the three categories. More than half (57%) of the articles generate a contribution of the qualitative or descriptive model whereas 24% of studies provide procedures and 19% of articles produce prototypes or tools.

**(5) Research Methods.** Classification of publications based on research methods has been often reported in mapping studies [59]. According to the mapping study guideline by Petersen *et al.* [59], the following research methodologies are frequently applied in SE mapping studies: *survey*, *case study*, *controlled experiment*, *action research*, *ethnography*, *simulation*, *prototyping*, and *mathematical analysis*. The description of these methodologies is taken from the Wieringa *et al.* [81], Wohlin *et al.* [83], and Easterbrook *et al.* [28]. In addition to these research methods, we also included *empirical methods* as a research methodology as defined by Basili *et al.* [14]. We also added *machine learning* as a separate method due to its prevalence in SO-based research. For the completeness of our categorization, we labeled research papers under *planning method* that only proposed a solution or procedure for their future research work. Finally, papers with multiple research methods (empirical, survey, prototyping, etc.) were labeled under the *mixed-methods* research. Figure 10a shows the distribution of research papers with different research methods. Most of the papers fall into the *empirical method* category, followed by the *machine learning* method category. For example, the paper "A Large-Scale Empirical Study on Linguistic Antipatterns Affecting APIs" [3] falls into the *empirical method* category, whereas the paper "DeepLocalize: Fault Localization for Deep Neural Networks" is within the *machine learning* category. The lowest number of papers are from the *survey* and *planning* categories. Examples of papers with the *survey* and *planning* methods are "Finding help with programming errors: An exploratory study of novice software engineers' focus in stack overflow posts" [22] and "Context-Aware Software Documentation" [2]. The former work surveyed developers to understand how they focus on a SO post while learning, whereas the later paper proposed the elementary version of a proposed framework for generating fine-grained code comments.

**(6) Research Objective.** During the pilot review of the 44 research papers (Section 2.2), the first author formed a common pattern of research objectives, which were finalized after a discussion with the second and the fourth author. These objectives (i.e., attributes for the research objective facet) include *tool development*, *topic analysis*, *post improvement*, *user behavior analysis*, *profile analysis*, *post* *analysis*, and *code analysis*. Papers that aimed to develop tools leveraging SO data are within the ***tool development*** category—e.g., "DiffTech: Differencing Similar Technologies from Crowd-Scale Comparison Discussions" [78]. Research papers that focused on understanding developers' knowledge on a certain topic are in the ***topic analysis*** category—e.g., "What do developers know about machine learning: a study of ml discussions on stackoverflow" [12]. *Post improvement* papers aim to improve the quality of the SO posts for reaching the appropriate audience—e.g., "Generating question titles for stack overflow from mined code snippets" [31]. The *user behavior analysis* category contains papers that investigate how SO users behave in different scenarios. For example, the paper "Stack Overflow Badges and User Behavior: An Econometric Approach" [46] shows that some users reduce their contributions to SO after receiving some particular group of badges. The *profile analysis* category includes papers that investigate different aspects of the SO users (e.g., age, gender, and expertise). For example, the paper "Gender differences in participation and reward on Stack Overflow" [47] shows that on average men are rewarded more than women for answering questions—a probable demonstration of gender bias in SO. The *post analysis* category contains paper that analyzes SO posts (e.g., tags, content, scores, etc.) without suggesting any improvement technique—e.g, "Improving Low-Quality Stack Overflow Post Detection" [61].

Figure 10b shows the distribution of the SO studies in all the discussed categories. Clearly, SO has been used to develop a significant number of practical tools—117 of all the studied research papers leveraged SO to develop software tools. This is encouraging because the practitioners' community reported the need for different research-based tools for integrating them within the software development processes [45, 64, 23, 75].

**(7) SO Content Usage.** Since SO contains a large volume of textual and software engineering content, there could be multi-faceted use of these contents in the literature. To identify how this content is used, we categorized the literature into three attributes - *content analytics*, *content extraction*, and *content quality*. *Content analytics* refers to the papers which focus on the analysis of the contents and perform various empirical, quantitative, or qualitative analyses of SO posts, comments, or tags. For example, the article "What are developers talking about? An analysis of topics and trends in stack overflow" [13] analyses SO posts to identify the broad technical topics and trends discussed in SO.

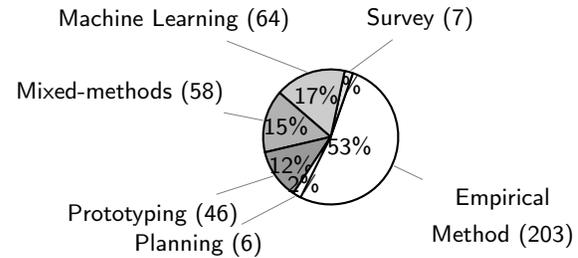

(a) Distribution research methods.

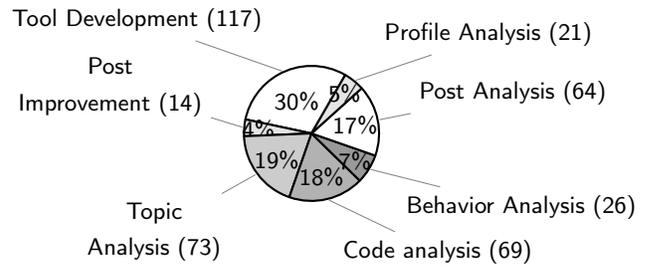

(b) Distribution of research objectives.

**Figure 10:** Distribution of facets research methods and objectives.

Some research extracts SO contents to use in training machine learning models or developing new tools/techniques since SO is a massive source of natural language corpus; these papers fall into the attribute of *content extraction*. The article "Enhancing python compiler error messages via stack" [74] extract SO contents to provide improved compiler errors.

Finally, there are a few research works that analyze the quality of SO contents. Such *quality analysis* may target the question-answering quality or sometimes can also focus on the quality of machine learning models for natural language processing trained on the SO contents. For example, "An empirical study of obsolete answers on Stack Overflow" [86] finds out the reasons for obsolete answers in SO and shows that more than half of the obsolete answers are already obsolete when they are posted, indicating the concerning quality of some SO answers.

Figure 11a shows the paper distribution based on the SO content usage. Two-thirds of the articles analyze SO contents whereas almost one-third of the articles extract SO contents and use that for external purposes.

**(8) Beneficiary.** As SO is a developer-focused question-and-answer platform, a significant number of research on SO

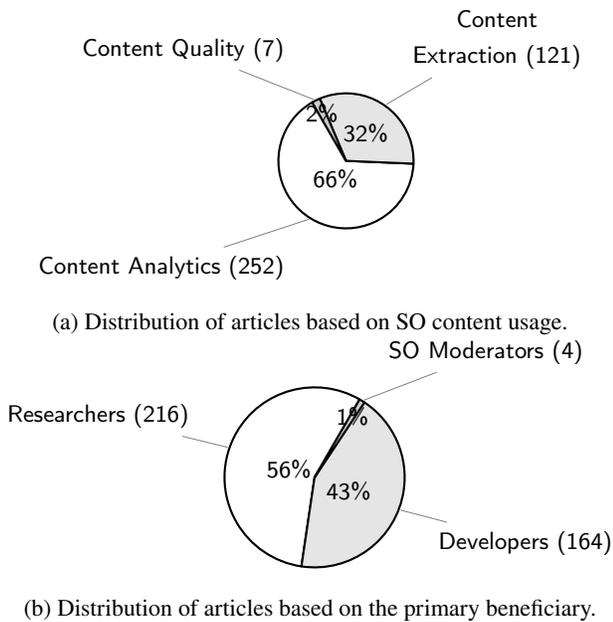

(a) Distribution of articles based on SO content usage.

(b) Distribution of articles based on the primary beneficiary.

**Figure 11:** Distribution of facets content usage and beneficiary

can also benefit developers besides researchers in academia. Moreover, some research can help organizations and SO moderators as well. For example, "Python Coding Style Compliance on Stack Overflow" [9] has a primary audience of researchers in the academia, whereas developers would get major benefits from articles like "Automatic tag recommendation for software development video tutorials" [55]. Few articles can be directly used by SO developers or moderators, for example, "Mining Duplicate Questions in Stack Overflow" [4]. Figure 11b shows that both academia and industry are very close primary beneficiaries for SO-based studies. 56% of studies primarily benefited researchers whereas 43% of studies contributed to the developers mostly.

**(9) SO Data Field.** SO contains a handful of data fields around which different research can be built. We identified the prominent data field which is used in the research methodology as a data source. Such data fields can be *post*, *question*, *answer*, *comment*, *code snippet*, *tag*, *internal/external link*, *user profile*, *user reputation*, *search log (from a web server)*, and *revision history*. For example, "AnswerBot: an answer summary generation tool based on stack overflow" [20] mined SO *answers* for generating the summary, whereas "Snakes in paradise?: Insecure python-related coding practices in stack overflow" [63] analyzed *code snippets* of SO posts to identify security vulnerabilities. If a paper uses any of at least two items of questions, answers, and comments, we labeled it as a '*post*' attribute in general.

As Figure 12a shows that more than two-thirds of the studies used SO *posts* as their data source which means that those studies used more than one particular data (such as *answers* and *code snippets*). 52 articles (14%) used only *code snippets* present in either the questions or answers of the SO posts. All other data sources (such as a *question*, *answer*, *tags*) are used less than 10%. Other data sources such as *user reputation*, *user profile*, *URL*, and *comment* constituted less than 10 articles in total and hence were grouped together as Other data fields in Figure 12a.

**(10) Data Movement Type.** Since SO is a massive storage of lexical and engineering data, often this data is used for the improvement of internal or external systems. For example, "Automated construction of a software-specific word similarity database" [75] has used SO posts for developing a technological vocabulary such as WordNet [51] that can be used in external machine learning systems. Under the Data Movement facet, we have labeled such articles with the *internal to external* attribute since the study is utilizing internal (SO) data for usage in external domains.

SO data can also be used to improve the overall SO system. For example, identifying duplicate posts and raising flags of duplication which can be used internally to the SO. "Generating Question Titles for Stack Overflow from Mined Code Snippets" [31] provided a mechanism to suggest question titles for SO posts based on the code snippet of the post. Such articles are using SO data to improve or analyze SO systems and hence labeled with the *internal to internal* attribute.

Lastly, there can be another use case of applying external SE techniques or ML models and datasets to improve SO posts such as suggesting appropriate tags for a post. For example, "Augmenting Stack Overflow with API Usage Patterns Mined from GitHub" [65] developed a Chrome extension, ExampleCheck that detects API usage violations in SO posts using API usage patterns mined from 380K external GitHub projects. Such articles are labeled under the *external to internal* attribute for the data movement facet. As Figure 12b shows most articles (44%) use SO data for analyzing or improving SO systems. 39% of papers attempt to use the SO data for application into external problem domains.

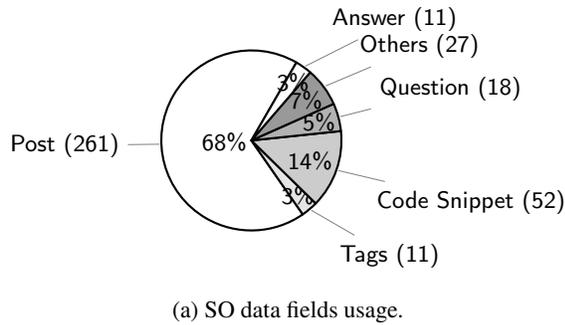

(a) SO data fields usage.

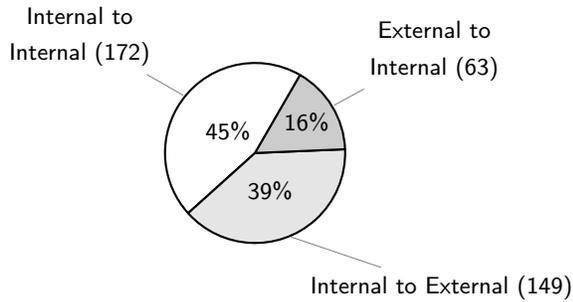

(b) SO data movement between inside and outside SO

**Figure 12:** Distribution of articles based on facets of SO data and data movement type

---

**RQ2 Summary:** We have developed a database of 384 SO-based research papers by dividing them into 10 facets comprising 63 attributes. We believe this catalog will help researchers find relevant SO-based research papers quickly. In general, SO-based research focuses a lot on tool development. In the case of SDLC, half of the research papers focus on the software development phase. Validation research (without industrial evaluation) is the most common research type (81%) and more than half of studies generate qualitative or descriptive models (57%) as research contributions. Also, empirical methods are the most prominent (53%) research method for SO-based studies.

---

## 5. (RQ3) What SE Domains are Studied in SO-based SE Research and what are their Impacts?

It is beyond the scope of our study to discuss all 10 facets and all of their attributes (63 in total) in detail. We, however, believe that the SE domains facet could be most useful to the research community and deserves a more in-depth analysis. Therefore, in this RQ, we describe the SE domain facet with all of its attributes (i.e., all the domains) in detail. In general, we show the trend and impact of the published papers in different SE domains and sub-domains (sub-attributes).

### 5.1. SE Domain Taxonomy

As we mentioned before, after assigning each paper to a specific domain, a total of 18 unique SE domains were found. Table 8 shows all the domains and the total counts of publications and average citations in each of those domains. The *API Design and Evolution* domain has the most papers (60), whereas the *Requirement Engineering* and *Program Repair*, despite their high average citations, have the lowest number of papers (2).

We also observed that we can not only divide these domains into more fine-grained sub-domains but also can merge multiple domains into six larger categories. Figure 13 shows the result after merging (the sub-domains are presented in Section 5.2). We left the domain *Software Analytics* in its own category as the domain itself is vast and contained almost thirty sub-domains under it. The rest of the categories are AI for SE (Recommender Systems, and Machine Learning with/for SE), Software Reuse and Assistance (API Design and Evolution, Tools and Environments), Human Aspects of SE (Social Aspects of SE, Crowd-based SE, Human-Computer Interaction), Program Analysis (Program Synthesis, Program Comprehension, Program Repair), Development Process (Testing and Analysis, Software Architecture and Design, Requirements Engineering), and Safety (Privacy and Security, Ethics in SE).

Among all the SE domain categories, AI for SE has the highest number of 111 papers which is almost one-third of all papers. AI for SE is followed by Software Reuse and Assistance category which has 75 papers published. These two categories account for nearly half of all research publications. Human Aspects of SE is another major research category (70 papers) that evolves around developers' interaction related to SO. Among all the categories, the Safety category has the least amount of papers (6%).

The categories and domain-wise publication count, percentage, and yearly publication trend are shown in Figure 14. AI for SE has constant growth in publication count every year though there was a sudden peak in publication a few years ago. Software Reuse and Assistance is another consistent SE research area that has seen publication growth since the beginning. On the contrary, Human Aspects of SE had significant publication during the early years of SO, and after a middle dip, there is again some traction in recent years. Software Analytics had only a sudden peak in

**Table 8**

SE domain wise publication count and per publication average citation. Black = publication, Grey = Average citation

| SE Domain | Publication Count | Average Citation |
|---|---|---|
| API Design and Evolution | 60 | 35 |
| Recommender Systems | 59 | 33 |
| Machine Learning with and for SE | 52 | 39 |
| Software Analytics | 45 | 55 |
| Social aspects of software engineering | 40 | 44 |
| crowd-based software engineering | 26 | 39 |
| Programming Languages | 20 | 21 |
| Privacy and Security | 19 | 26 |
| Tools and Environments | 15 | 29 |
| Testing and Analysis | 11 | 31 |
| Release engineering and DevOps | 11 | 24 |
| Software Architecture and Design | 6 | 6 |
| Program synthesis | 5 | 32 |
| Human-computer interaction | 4 | 19 |
| Ethics in Software Engineering | 4 | 39 |
| Program comprehension | 3 | 19 |
| Requirements Engineering | 2 | 36 |
| Program repair | 2 | 45 |

a couple of years, other than that this domain is mostly flat. Program Analysis has almost consistent paper publication except for the first few years. Both Development Process and Safety categories have signification publications only in recent years.

## 5.2. SE Domain-wise Studies

We now discuss all the major domains separately with their sub-domains. The domains that have very few publications are discussed together in Section 5.2.12. Our objective is to find which domains (and sub-domains in a specific domain) are most prevalent and impactful in SO-based SE research. For this analysis, we use a set of metrics that we describe as follows.

**(i) Overall Impact Factor (OI).** Inspired by the Journal Impact Factor [32], we calculate the OI with Equation 1. Here, OI is *Overall Domain Impact Factor*, PC is *Total Publication Count in a Domain*, and CD is *Citation Density*.

$$OI = \frac{PC + CD}{2} \qquad (1)$$

The citation density [33] is calculated as the average citation per publication in a research domain as defined in Equation 2 (TC is *Total Citation of Publications in a Domain*).

$$CD = \frac{TC}{PC} \qquad (2)$$

**(ii) Recent Impact Factor (RI).** RI considers the publications from the years 2019, 2020, and 2021. We have observed that these three years together contribute to a total of 196 publications which is more than half of all studied publications (384). Equation 3 for the RI is similar to OI except that it considers publications and citations for these three years only. Here, RPC is *Recent Publication Count in a Domain*, and RCD is *Recent Citation Density*.

$$RI = \frac{RPC + RCD}{2} \qquad (3)$$

**(iii) Impact Ratio (%I).** We also calculated the impact ratio (%I) of each domain as $OI/RI$.

**(iv) Potential Impact Factor (PI).** Finally, we noticed that some domains has more publications in recent years (since 2019) compared to other domains. To reflect this distinction among multiple domains, we have calculated a potential impact factor (PI) where the percentage of recent publication count is multiplied by the recent impact factor as described in Equation 4 (PC is *Total Publication Count in a Domain* and RPC is *Recent Publication Count in a Domain*).

$$PI = \frac{RPC}{PC} x RI \qquad (4)$$

**Productivity Type.** We also calculated the type of productivity for a given sub-domain using their citation density and publication counts. For each domain, we calculated the median publication count and median citation density among all the sub-domains which has more than one publication. Then we divided all the sub-domains among four quadrants as shown in Figure 15. The four quadrants are defined as followed.

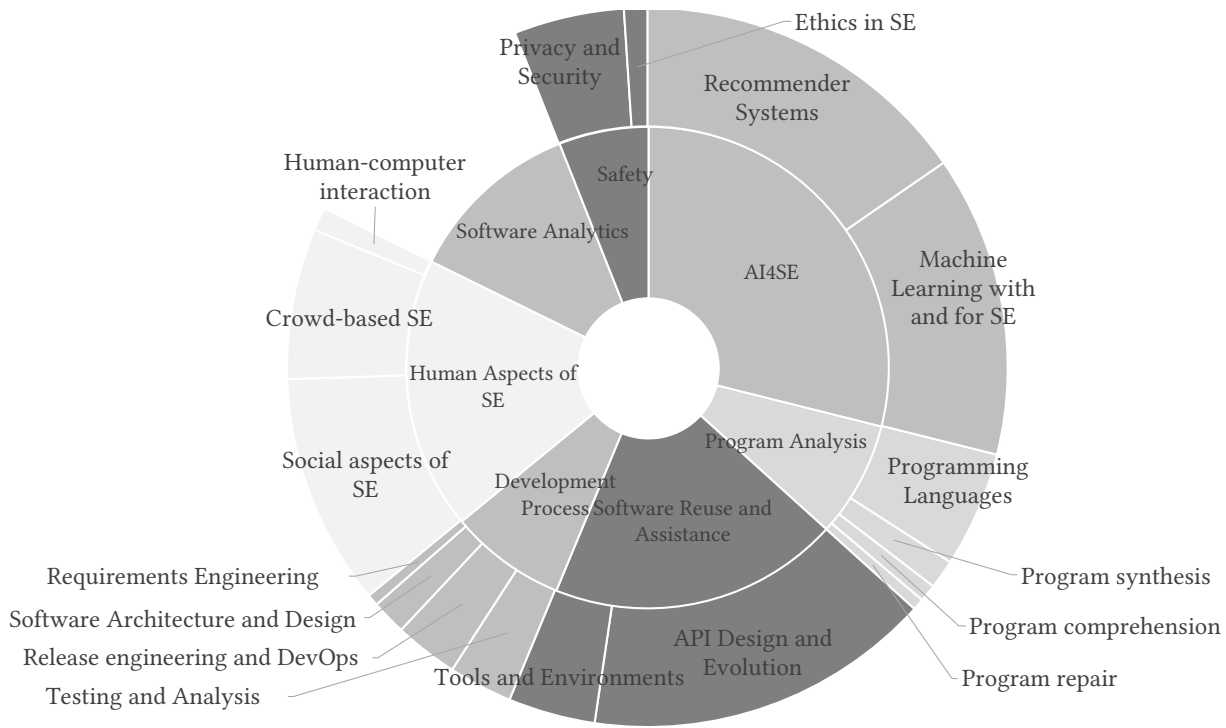

**Figure 13:** Research distribution by SE Domain categories

- *Fertile Zone (F)*. A sub-domain has more citation density than the median citation density and more publication than the median publication of other sub-domains. Sub-domains in this area are considered fertile because there is much active research with high citations.

- *Crowded Zone (C)*. A sub-domain has less citation density than the median citation density but more publications than the median publication. Sub-domains in this zone are considered crowded because there is much active research but on average this domain is not attracting the other research community significantly (i.e., fewer citations).

- *Green Zone (G)*. Citation density is higher than the median citation density but there are fewer publications than the median publication count. We can say that such sub-domains have less research despite their potential, and therefore, should be studied more in the future.

- *Barren Zone (B)*. Citation density and the publication count of a sub-domain both are less than the median citation and publication count.

Figure 16 shows all 18 domains according to the four productivity zones. Researchers can focus on domains in the fertile and green zones (e.g., Software Analytics, Program Repair, etc.) for producing impactful research. Now, we analyze the sub-domains of each SE domain and identify their productivity zone in addition to their trend and different impact factors.

### 5.2.1. Recommender Systems

Recommender Systems ( ▪▪▪▪▪▪▪▪ ) has an overall impact of 46 and a recent impact of 11 (20% of overall impact) meaning that there has been a relatively lower amount of research in recent years (Table 9). IDE Plugin sub-domain ( ▌▌▪▪▪▪▪▪ ) falls into the fertile zone and has the most overall impact (OI=38) though its recent impact is low (RI=5.5). The highest number of publications comes (17) from Search Optimization ( ▪▪▪▪▌▌▌▌ ) which has consistent growth of publications over the years and has recent active research in the area, but it has one of the lowest citation density

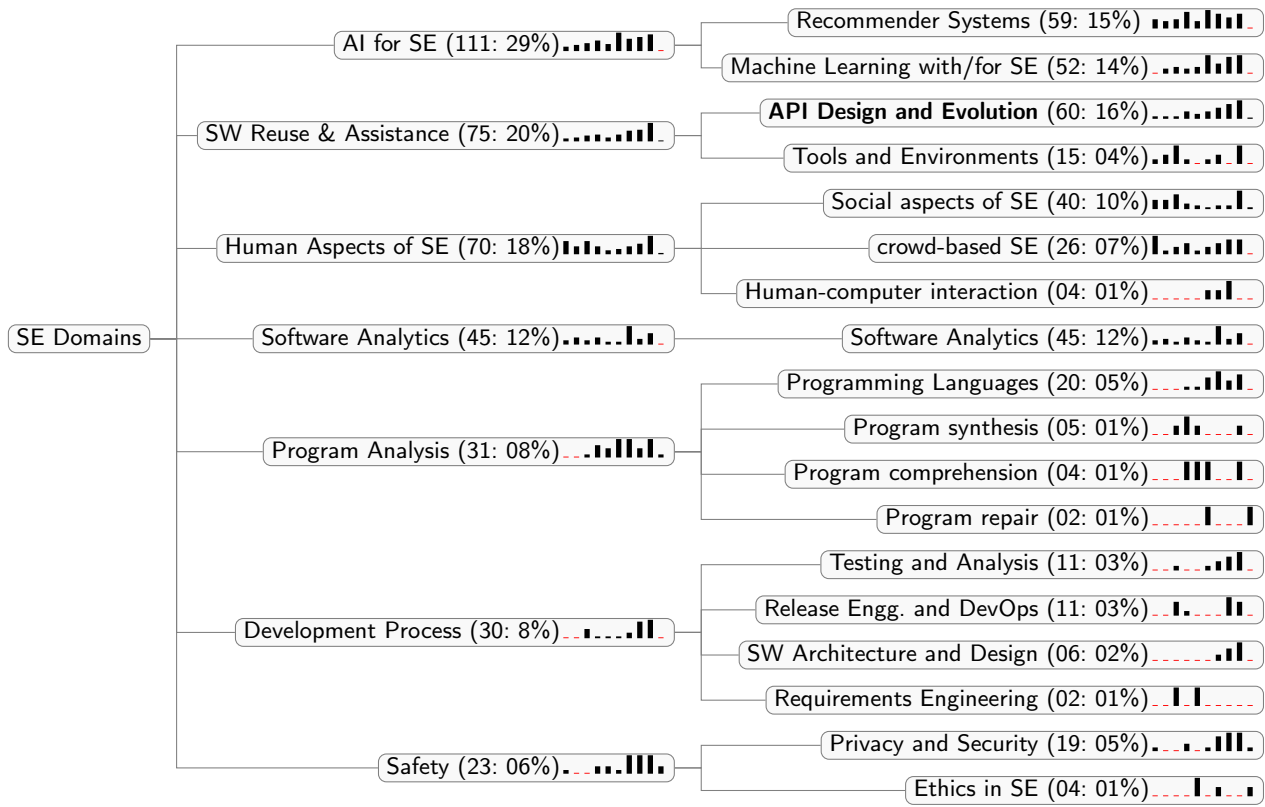

**Figure 14:** Taxonomy and yearly trend of SE domain-specific publications. Among all the SE domains, API Design and Evolution has the highest number of SO-based papers. The inline bar chart represents the publication trend for the last 10 years.

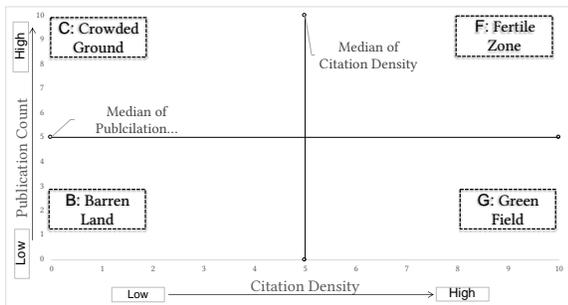

**Figure 15:** Domain or Sub-domain Productivity Chart

(18) among the sub-domains of recommender systems. It, therefore, falls into the crowded zone. Due to its low number of publications but a very high citation density, Video Extraction falls into the green zone—i.e., future research should focus more on this sub-domain.

### 5.2.2. Machine Learning with and for SE

Research publications in machine learning ( ) techniques around SO have grown consistently high under different sub-domains with an overall impact of 45.5 as shown in Table 10. ML Model/Data Preparation ( ) is a crowded domain that has the highest number of publications (18) but has a relatively lower citation density. Sentiment Analysis, ML Library Evaluation, Duplicate Detection, Post Summarization, and LDA Configuration all these domains fall into the fertile zone. Unlike many other domains, none of the sub-domains of this domain fall into the barren zone.

### 5.2.3. API Design and Evolution

As shown in Table 11, in API Design and Evolution, API Documentation is clearly the most fertile sub-domain with continuously increasing yearly publications ( ). API version detection ( ) and API evolution ( ) both lie inside the green field, because of their high citation density and low number of publications. Although API Issues ( ) and API Comparison ( ) both are crowded sub-domains, API comparison has much more recent active research compared to its beginning.

### 5.2.4. Tools and Environments

Tools and Environments ( ) domain has 15 publications under 9 sub-domains and have 29 citation density as shown in Table 12. This domain does not show

**Table 9**

Productivity, trend, and different impact factors of Recommendation Systems and its sub-domains (OI-overall impact, RI-recent impact, %I impact change rate, PI-potential impact, C-Crowded Zone, G-Green Zone, F-Fertile Zone, B-Barren Zone.

| (Sub)Domain | | Pub | Cite | Trend | OI | RI | %I | PI |
|---|---|---|---|---|---|---|---|---|
| **Recommender Systems** | | 59 | 33 | | 46 | 11 | 0.2 | 4.1 |
| **C** | Search Optimization | 17 | 18 | | 17.5 | 7 | 0.4 | 3.3 |
| **F** | Post Tag Recommendation | 9 | 58 | | 33.5 | 2 | 0.1 | 0.2 |
| **F** | IDE Plugin | 9 | 67 | | 38 | 5.5 | 0.1 | 1.8 |
| **C** | Query Reformulation | 7 | 20 | | 13.5 | 6.5 | 0.5 | 2.8 |
| **B** | Post Summarization | 4 | 17 | | 10.5 | 6.5 | 0.6 | 4.9 |
| **G** | Video Extraction | 3 | 46 | | 24.5 | 15.5 | 0.6 | 5.2 |
| **G** | Search Translation | 3 | 31 | | 17 | 0 | 0 | 0 |
| **B** | Duplicate Detection | 3 | 9 | | 6 | 1 | 0.2 | 0.3 |
| Other Sub-domains | | 4 | 12 | | 8 | 1 | 0.1 | 0.5 |

**Table 10**

Productivity, trend, and different impact factors of ML with and for SE and its sub-domains (OI-overall impact, RI-recent impact, %I impact change rate, PI-potential impact, C-Crowded Zone, G-Green Zone, F-Fertile Zone, B-Barren Zone.)

| (Sub)Domain | | Pub | Cite | Trend | OI | RI | %I | PI |
|---|---|---|---|---|---|---|---|---|
| **Machine Learning with/for SE** | | 52 | 39 | | 45.5 | 13.5 | 0.3 | 7 |
| **C** | ML Model/Data Preparation | 18 | 42 | | 30 | 9.5 | 0.3 | 3.2 |
| **F** | Sentiment Analysis | 5 | 53 | | 29 | 9.5 | 0.3 | 3.8 |
| **F** | ML Library Evaluation | 4 | 49 | | 26.5 | 3.5 | 0.1 | 1.8 |
| **F** | Duplicate Detection | 2 | 56 | | 29 | 0 | 0 | 0 |
| **F** | Post summarization | 2 | 43 | | 22.5 | 2.5 | 0.1 | 1.3 |
| **C** | Edit Suggestion | 2 | 0 | | 1 | 1 | 1 | 1 |
| **C** | Post Title Recommendation | 2 | 8 | | 5 | 5 | 1 | 5 |
| **C** | Knowledge Graph | 2 | 28 | | 15 | 4 | 0.3 | 2 |
| **F** | LDA Configuration | 2 | 89 | | 45.5 | 13 | 0.3 | 6.5 |
| **C** | Machine Learning | 2 | 11 | | 6.5 | 6.5 | 1 | 6.5 |
| Others Sub-domains | | 11 | 31 | | 21 | 4 | 0.2 | 2.9 |

**Table 11**

Productivity, trend, and different impact factors of API Design and Evolution its sub-domains (OI-overall impact, RI-recent impact, %I impact change rate, PI-potential impact, C-Crowded Zone, G-Green Zone, F-Fertile Zone, B-Barren Zone.)

| (Sub)Domain | | Pub | Cite | Trend | OI | RI | %I | PI |
|---|---|---|---|---|---|---|---|---|
| **API Design and Evolution** | | 60 | 35 | | 47.5 | 18 | 0.4 | 10.8 |
| **F** | API Documentation | 28 | 45 | | 36.5 | 15 | 0.4 | 9.1 |
| **C** | API Recommendation | 9 | 28 | | 18.5 | 6.5 | 0.4 | 4.3 |
| **C** | API Issues | 8 | 25 | | 16.5 | 7.5 | 0.5 | 3.8 |
| **C** | API Comparison | 7 | 18 | | 12.5 | 5 | 0.4 | 3.6 |
| **G** | API Evolution | 3 | 59 | | 31 | 3.5 | 0.1 | 2.3 |
| **B** | Code Usage | 2 | 17 | | 9.5 | 0 | 0 | 0 |
| **G** | API Version Detection | 2 | 31 | | 16.5 | 1.5 | 0.1 | 0.8 |
| Others Sub-domains | | 1 | 3 | | 2 | 0.5 | 0.3 | 0.5 |

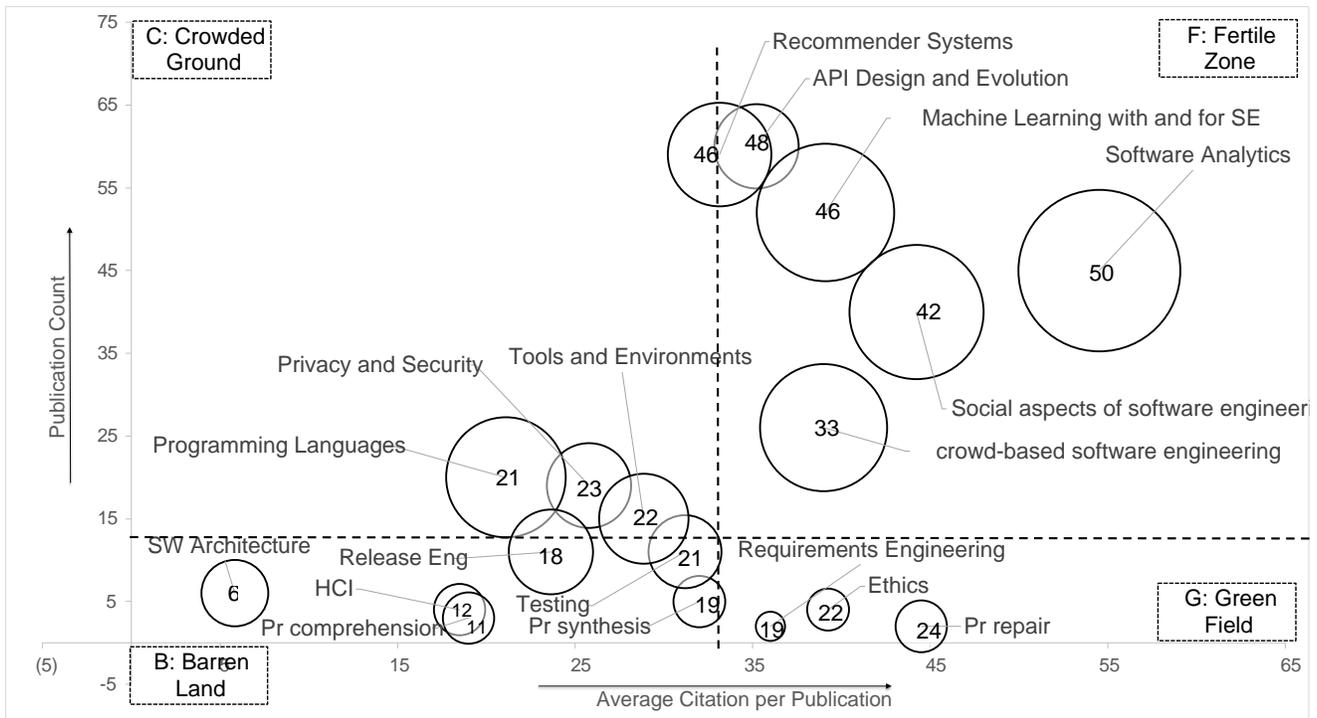

**Figure 16:** Average Citation and Publication Count for SE domains. The size of the bubble denotes the number of sub-domains under each domain. The number inside the bubble denotes the overall impact factor of the domain.

**Table 12**

Productivity, trend, and different impact factors of Tools and Environment and its sub-domains (OI-overall impact, RI-recent impact, %I impact change rate, PI-potential impact, C-Crowded Zone, G-Green Zone, F-Fertile Zone, B-Barren Zone.)

| | (Sub)Domain | Pub | Cite | Trend | OI | RI | %I | PI |
|---|---|---|---|---|---|---|---|---|
| | **Tools and Environments** | 15 | 29 | ▪▪▌▪▪▪▌ | 22 | 3 | 0.1 | 1.2 |
| **C** | IDE Plugin | 5 | 38 | ▪▌▪▌ | 21.5 | 0 | 0 | 0 |
| **C** | Execution Environment | 2 | 14 | ▪▌▌ | 8 | 8 | 1 | 8 |
| **F** | SO Torrent | 2 | 69 | ▐▐ | 35.5 | 25.5 | 0.7 | 12.8 |
| | Others Sub-domains | 6 | 13 | ▪▌▌ | 9.5 | 1.5 | 0.2 | 0.8 |

any consistency in publication trend, as there was initial growth followed by a downturn and few recent publications. Three sub-domains IDE Plugin (▪▌▌▌▁▁▁▁), Execution Environment (▁▁▁▁▁▌▁▌), and SO Torrent (▁▁▁▁▐▐▁▁) have more than one publication. Though SO Torrent has the highest impact factor (35.5) and falls into the fertile zone, it had no publication in the last three years. Another high impact (OI=21.5) sub-domain IDE Plugin also had publications only in the first four years and no publications generated thereafter. The other six sub-domains (Post summarization, Package Manager, Debug resolution, SQL Recommendation, Browser Plugin, and Knowledge Graph) constitute 13 citation density only.

### 5.2.5. Social aspects of SE

As shown in Table 13 Social Aspects of SE (▪▪▌▪▪▪▪▌) has been an active research domain since the beginning of SO-based research. Though there was a downturn in publications since 2014, there has been active research in the last two years. SO users' Reputation evaluation is the most impactful sub-domain (▐▪▌▪▌▁, OI=31.5), followed by users' Expertise (▪▌▁▁▁▁▁▁, OI=22), Answering Time (▁▪▌▁▪▁▁▁, OI=21.5), and Gender Bias (▁▁▁▁▪▌▁, OI=20.5). All these four sub-domains also fall under the fertile zone because of high citations and publications. Two active research areas regarding Community Engagement (▐▌▪▌▁▪▌) and Link Sharing (▁▁▁▁▁▁▌▌) fall into the crowded zone as they fail to generate enough citations. SO

**Table 13**

Productivity, trend, and different impact factors of Social aspects and its sub-domains (OI-overall impact, RI-recent impact, %I impact change rate, PI-potential impact, C-Crowded Zone, G-Green Zone, F-Fertile Zone, B-Barren Zone.)

| (Sub)Domain | | Pub | Cite | Trend | OI | RI | %I | PI |
|---|---|---|---|---|---|---|---|---|
| **Social aspects of SE** | | 40 | 44 | | 42 | 7 | 0.2 | 2.5 |
| F | Expertise | 4 | 40 | | 22 | 6.5 | 0.3 | 1.6 |
| C | Community Engagement | 4 | 12 | | 8 | 0.5 | 0.1 | 0.1 |
| F | Answering Time | 4 | 39 | | 21.5 | 0 | 0 | 0 |
| F | Gender Bias | 4 | 37 | | 20.5 | 10.5 | 0.5 | 7.9 |
| F | Reputation | 3 | 60 | | 31.5 | 0.5 | 0 | 0.2 |
| C | Link sharing | 3 | 17 | | 10 | 1 | 0.1 | 0.3 |
| B | Resume Generation | 2 | 7 | | 4.5 | 4.5 | 1 | 4.5 |
| G | Badges | 2 | 29 | | 15.5 | 18.5 | 1.2 | 9.3 |
| B | Answer Accceptance | 2 | 22 | | 12 | 0 | 0 | 0 |
| B | Covid-19 | 2 | 2 | | 2 | 2 | 1 | 2 |
| | Others Sub-domains | 10 | 91 | | 50.5 | 1 | 0 | 0.2 |

Badges ( ) is the only sub-domain that falls inside the green field and also has a relatively higher recent impact than overall impact (OI=15.5, RI=18.5). Sub-domains regarding Resume Generation (OI=4.5), Answer Acceptance (OI=12) and Covid-19 (OI=2) all fall under barren land because of their low citation and publication. Interestingly, there are 10 other sub-domains having one publication each but their citation density is 91, and their overall impact is 50.5 which is higher than any other single sub-domain. Three research papers on Questions Types (citation 452), Personality Analysis (citation 154), and Post Quality Detection (citation 130) have contributed to this high impact.

### 5.2.6. Crowd-based SE

Crowd-based SE ( ) domain has moderate publication (26) and citation (39) resulting in an impact factor of 39 (Table 14). It had an initial surge in publications, and in the last few years, the publication count increased steadily. Its Answering Time ( OI=12) sub-domain has the most number of publications, but because of lower citation density, the sub-domain falls into the crowded zone. SO Usefulness ( , OI=26.5), Gamification ( OI=15.5), and Link Sharing ( OI=14.5) are the three most impactful sub-domains and fall into the fertile zone. There are other 11 sub-domains each consisting of only one publication and have a citation density of 59.

### 5.2.7. Software Analytics

Software Analytics ( ) is a high impact (OI=50) domain with 45 publications and 55 citation density. As shown in Table 15, Code Cloning ( ) is a crowded domain (because of relatively higher publications but lower citations compared to other sub-domains). It has very recent active research with an RI score of 16. SO Discussion Topics ( ) and Mobile Development ( ) are two high-impact sub-domains (OI 114.5 and 67 respectively) putting them into the fertile zone, although none of them have active research publications in recent years. Noticeably, Software Energy Consumption falls into the green zone because of its high citation but a low number of publications, implying the usefulness of this domain to the community.

### 5.2.8. Programming Languages

Programming Language ( ) had a consistent publication trend in the last five years with 20 publications, and 21 citation density resulting in an overall impact of 20.5 (Table 16). Though there are a total of 16 sub-domains, only three sub-domains FQN Searching ( ), Code Quality ( ), and Language Detection ( ) have more than one publication. Among them, FQN Searching falls into the fertile zone and the other two fall into the crowded ground.

### 5.2.9. Testing and Analysis

Testing and Analysis ( ) has a very strong recency trend in publications in the last five years (Table 17) with 5 sub-domains generating 11 publications and 31

**Table 14**

Productivity, trend, and different impact factors of Crowd-based SE and its sub-domains (OI-overall impact, RI-recent impact, %I impact change rate, PI-potential impact, C-Crowded Zone, G-Green Zone, F-Fertile Zone, B-Barren Zone.)

| (Sub)Domain | | Pub | Cite | Trend | OI | RI | %I | PI |
|---|---|---|---|---|---|---|---|---|
| **Crowd-based** SE | | 26 | 39 | | 32.5 | 5.5 | 0.2 | 2.3 |
| C | Answering Time | 3 | 21 | | 12 | 1.5 | 0.1 | 0.5 |
| F | Gamification | 2 | 29 | | 15.5 | 6.5 | 0.4 | 3.3 |
| C | Answer Triaging | 2 | 22 | | 12 | 0 | 0 | 0 |
| F | Link sharing | 2 | 27 | | 14.5 | 1 | 0.1 | 0.5 |
| F | SO Usefulness | 2 | 51 | | 26.5 | 0 | 0 | 0 |
| C | Comment Analysis | 2 | 18 | | 10 | 10 | 1 | 10 |
| C | Edit History | 2 | 4 | | 3 | 3 | 1 | 3 |
| Others Sub-domains | | 11 | 59 | | 35 | 2 | 0.1 | 0.7 |

**Table 15**

Productivity, trend, and different impact factors of Software Analytics and its sub-domains (OI-overall impact, RI-recent impact, %I impact change rate, PI-potential impact, C-Crowded Zone, G-Green Zone, F-Fertile Zone, B-Barren Zone.)

| (Sub)Domain | | Pub | Cite | Trend | OI | RI | %I | PI |
|---|---|---|---|---|---|---|---|---|
| **Software Analytics** | | 45 | 55 | | 50 | 12.5 | 0.3 | 6.9 |
| C | Code Cloning | 7 | 31 | | 19 | 16 | 0.8 | 13.7 |
| C | Code Usage | 4 | 76 | | 40 | 9.5 | 0.2 | 2.4 |
| F | Mobile Development | 3 | 131 | | 67 | 0 | 0 | 0 |
| F | SO Discussion Topics | 3 | 226 | | 114.5 | 4.5 | 0 | 1.5 |
| B | Technology Landscape | 2 | 43 | | 22.5 | 0 | 0 | 0 |
| B | Code Quality | 2 | 22 | | 12 | 2 | 0.2 | 1 |
| G | Energy Consumption | 2 | 116 | | 59 | 0 | 0 | 0 |
| Others Sub-domains | | 22 | 23 | | 22.5 | 8 | 0.4 | 5.8 |

citation density. Deep Learning Bug Analysis (⣀⣀⣀⣀⣀ ) sub-domain stands out among all the sub-domains for higher publication and citation density resulting in superior impact. While it falls in the fertile zone, the other two sub-domains Bug Analysis (⣀⣀⣀⣀⣀ ) and Deep Learning Fault Detection (⣀⣀⣀⣀⣀ ) both are crowded areas as they have lower citations. However, all three sub-domains have a very good recency impact indicating that overall such analysis is getting traction for using SO.

### 5.2.10. Release engineering and DevOps

Release and Engineering (⣀⣀⣀⣀⣀ ) domain has 11 publications under 8 sub-domains where only two domains have more than two publications as shown in Table 18. Community Support using SO was an early research area

**Table 16**

Productivity, trend, and different impact factors of Programming Language and its sub-domains (OI-overall impact, RI-recent impact, %I impact change rate, PI-potential impact, C-Crowded Zone, G-Green Zone, F-Fertile Zone, B-Barren Zone.)

| (Sub)Domain | | Pub | Cite | Trend | OI | RI | %I | PI |
|---|---|---|---|---|---|---|---|---|
| **Programming Languages** | | 20 | 21 | | 20.5 | 7 | 0.3 | 4.9 |
| F | FQN Searching | 3 | 13 | | 8 | 3 | 0.4 | 2 |
| C | Code Quality | 2 | 8 | | 5 | 5 | 1 | 5 |
| C | Language Detection | 2 | 7 | | 4.5 | 4.5 | 1 | 4.5 |
| Others Sub-domains | | 13 | 27 | | 20 | 4 | 0.2 | 2.5 |

**Table 17**

Productivity, trend, and different impact factors of Testing and Analysis and its sub-domains (OI-overall impact, RI-recent impact, %I impact change rate, PI-potential impact, C-Crowded Zone, G-Green Zone, F-Fertile Zone, B-Barren Zone.)

| (Sub)Domain | | Pub | Cite | Trend | OI | RI | %I | PI |
|---|---|---|---|---|---|---|---|---|
| **Testing and Analysis** | | 11 | 31 | | 21 | 4.5 | 0.2 | 3.7 |
| **F** | Deep Learning Bug Analysis | 5 | 55 | | 30 | 20 | 0.7 | 16 |
| **C** | Bug Analysis | 2 | 9 | | 5.5 | 5.5 | 1 | 5.5 |
| **C** | Deep Learning Fault Detection | 2 | 4 | | 3 | 3 | 1 | 3 |
| Others Sub-domains | | 2 | 23 | | 12.5 | 0.5 | 0 | 0.3 |

**Table 18**

Productivity, trend, and different impact factors of Release Engineering and DevOps and its sub-domains (OI-overall impact, RI-recent impact, %I impact change rate, PI-potential impact, C-Crowded Zone, G-Green Zone, F-Fertile Zone, B-Barren Zone.)

| (Sub)Domain | | Pub | Cite | Trend | OI | RI | %I | PI |
|---|---|---|---|---|---|---|---|---|
| **Release engineering and DevOps** | | 11 | 24 | | 17.5 | 3.5 | 0.2 | 2.2 |
| **F** | Community Support | 3 | 41 | | 22 | 0 | 0 | 0 |
| **B** | Deep Learning Deployment | 2 | 20 | | 11 | 11 | 1 | 11 |
| Others Sub-domains | | 6 | 16 | | 11 | 2.5 | 0.2 | 2.1 |

that falls under the fertile zone as it has higher citation density compared to other sub-domains. However, Deep Learning Deployment is a very recent research sub-domain have all publications in the last 3 years.

### 5.2.11. Privacy and Security

Table 19 shows that Privacy and Security had strong growth in recent years ( ) with 19 publications and a moderate 26 citation density resulting in an overall impact of 22.5. Out of total 8 sub-domains, only two sub-domains Code Vulnerability ( , OI=9.5), and Cryptography ( , OI=26.5) have more than two publications. Code vulnerability falls into the crowded zone for its lower citation density, whereas Cryptography falls into the green zone.

### 5.2.12. Other SE Domains

As shown in Table 20, seven other SE domains (Software Architecture and Design, Program Synthesis, Human-computer interaction, Program Comprehension, Ethics in Software Engineering, Requirements Engineering, and Program repair) each have less than ten publications based on SO. Among them, Software Architecture has constantly increasing trend in recent publications on software design patterns and system architectures such as microservice [11, 82]. There are a few sporadic publications regarding Ethics in Software Engineering that investigated license issues in

SO code snippets and unauthorized code cloning among SO and other code repository platforms [8, 10]. Few studies are conducted on human interaction, emotion, and accessibility issues in SO under Human-computer interaction domain [22, 54]. There are very few publications on program repair, synthesis, and comprehension domains to analyze or generate source code, and comments [87, 44].

### 5.3. Top Sub-domains Across All Domains

While we have analyzed the most active and impactful sub-domains for a given domain, it is also important to compare sub-domains across all 18 domains. To achieve meaningful insights, we have applied a few filters among the total 125 sub-domains. Firstly, we have excluded any sub-domain that has less than 4 publications. Secondly, we have filtered out any sub-domain which has less than 30% publications since 2019 on the ground that such sub-domains had low activity in recent years (on average, 50% of papers were published on or after 2019). This filtering approach gives us 16 sub-domains which are sorted by the Recent Impact factor and presented in Table 21. Clearly, Deep Learning Bug Analysis is the most productive SE sub-domain considering the Recent Impact factor, and the second most productive if citation density is considered. However, if the number of publications is also taken into consideration, API Documentation is the most productive.

**Table 19**

Productivity, trend, and different impact factors of Security and Privacy and its sub-domains (OI-overall impact, RI-recent impact, %I impact change rate, PI-potential impact, C-Crowded Zone, G-Green Zone, F-Fertile Zone, B-Barren Zone.)

| | (Sub)Domain | Pub | Cite | Trend | OI | RI | %I | PI |
|---|---|---|---|---|---|---|---|---|
| | **Privacy and Security** | 19 | 26 | ▁▁▁▂▁▃▅▅▅▁ | 22.5 | 7 | 0.3 | 5.2 |
| **C** | Code Vulnerability | 9 | 10 | ▁▁▁▁▁▅▅▃▁ | 9.5 | 9.5 | 1 | 9.5 |
| **G** | Cryptography | 4 | 49 | ▁▁▁▁▂▁▅▁ | 26.5 | 3 | 0.1 | 2.3 |
| | Others Sub-domains | 6 | 34 | ▅▁▁▃▁▅▅▅▅ | 20 | 1 | 0.1 | 0.3 |

**Table 20**

Productivity, trend, and different impact factors of all other SE Domains each of which has less than 10 publications, and their sub-domains (OI-overall impact, RI-recent impact, %I impact change rate, PI-potential impact, C-Crowded Zone, G-Green Zone, F-Fertile Zone, B-Barren Zone.)

| Domains with <10 publications | Pub | Cite | Trend | OI | RI | %I | PI |
|---|---|---|---|---|---|---|---|
| Software Architecture and Design | 6 | 6 | ▁▁▁▁▁▂▅▅ | 6 | 3 | 0.5 | 3 |
| Program synthesis | 5 | 32 | ▁▂▅▃▁▁▂▁ | 18.5 | 0.5 | 0 | 0.1 |
| Human-computer interaction | 4 | 19 | ▁▁▁▁▂▅▅▂ | 11.5 | 1.5 | 0.1 | 1.1 |
| Program comprehension | 3 | 19 | ▁▁▅▅▅▁▁ | 11 | 0 | 0 | 0 |
| Ethics in Software Engineering | 4 | 39 | ▁▁▁▃▁▅▁▂ | 21.5 | 0.5 | 0 | 0.1 |
| Requirements Engineering | 2 | 36 | ▁▃▅▅▁▁▁ | 19 | 0 | 0 | 0 |
| Program repair | 2 | 45 | ▁▁▁▃▅▁▅ | 23.5 | 0 | 0 | 0 |

**RQ3 Summary:** We developed an indexed database of 18 SE domains and 125 unique sub-attributes (i.e., sub-domains) under those domains and labeled each of the 384 SO-based SE research papers accordingly. We have shown which SE domains (and sub-domains) were most (or least) explored by the research communities, and where should future research focus to be more helpful and attractive to the community. The research community can use our labeling of the green and fertile zones to understand what SO-based research can be most fruitful in the near future.

## 6. Discussion

### 6.1. Recommendations

In this paper, we have produced an indexed database of 384 SO-based SE research papers. In general, we have established that SO is the most prominent Q&A site for software engineering research (RQ1), and it attracts researchers from diverse software engineering communities. Our database of papers consists of 10 facets and 63 attributes that a researcher should find helpful for finding relevant research papers for a given area or topic of interest (RQ2). We then focused on the SE domain facet and found a staggering 18 domains and 125 sub-domains that used SO content (e.g., questions, answers, comments, code, etc.) for producing high-quality research, including techniques, tools, and prototypes. Our productivity measurements across these domains and sub-domains would help a researcher to decide which topic of SO-based research can be impactful in the future.

**Recommendation:** Researchers can consider the domains and sub-domains in the green and fertile zones (presented in RQ3) for producing impactful SO-based SE research. For example, the most fertile is the Software Analytics domain (Figure 16), whereas Code Cloning and Code Usage are the most fertile sub-domains in that domain (Table 15). A researcher can also use Table 21 for directly assessing the most potential sub-domains, which would suggest that Deep Learning Bug Analysis, Code Cloning, and API Documentation are the top potential sub-domains for SO-based SE research.

We have also observed that SO content (i.e., data) is increasingly used to model and improve external systems. Figure 17 shows the trend of how SO content has been used to improve SO itself (internal to internal), to improve external systems (internal to external), and how external data (e.g., data from GitHub) was used to improve SO. According

**Table 21**

List of the most productive SE sub-domains (OI-overall impact, RI-recent impact, %I impact change rate, PI-potential impact, R-Pub= percentage of publications since 2019). Sub-domains are listed in descending order of their recent impact (RI).

| Sub-domain | Domain | Pub | Cite | OI | RI | %I | R-Pub | PI |
|---|---|---|---|---|---|---|---|---|
| Deep Learning Bug Analysis | Testing and Analysis | 5 | 55 | 30.0 | 20.0 | 0.7 | 0.8 | 16.0 |
| Code Cloning | Software Analytics | 7 | 31 | 19.0 | 16.0 | 0.8 | 0.86 | 13.7 |
| API Documentation | API Design and Evolution | 28 | 45 | 36.5 | 15.0 | 0.4 | 0.61 | 9.1 |
| Gender Bias | Social aspects of SE | 4 | 37 | 20.5 | 10.5 | 0.5 | 0.75 | 7.9 |
| ML Model/Data Preparation | Machine Learning with and for SE | 18 | 42 | 30.0 | 9.5 | 0.3 | 0.33 | 3.2 |
| Sentiment Analysis | Machine Learning with and for SE | 5 | 53 | 29.0 | 9.5 | 0.3 | 0.4 | 3.8 |
| Code Vulnerability | Privacy and Security | 9 | 10 | 9.5 | 9.5 | 1.0 | 1 | 9.5 |
| API Issues | API Design and Evolution | 8 | 25 | 16.5 | 7.5 | 0.5 | 0.5 | 3.8 |
| Search optimization | Recommender Systems | 17 | 18 | 17.5 | 7.0 | 0.4 | 0.47 | 3.3 |
| API Recommendation | API Design and Evolution | 9 | 28 | 18.5 | 6.5 | 0.4 | 0.67 | 4.3 |
| Query Reformulation | Recommender Systems | 7 | 20 | 13.5 | 6.5 | 0.5 | 0.43 | 2.8 |
| Post summarization | Recommender Systems | 4 | 17 | 10.5 | 6.5 | 0.6 | 0.75 | 4.9 |
| IDE Plugin | Recommender Systems | 9 | 67 | 38.0 | 5.5 | 0.1 | 0.33 | 1.8 |
| API Comparison | API Design and Evolution | 7 | 18 | 12.5 | 5.0 | 0.4 | 0.71 | 3.6 |
| ML Library Evaluation | Machine Learning with and for SE | 4 | 49 | 26.5 | 3.5 | 0.1 | 0.5 | 1.8 |
| Cryptography | Privacy and Security | 4 | 49 | 26.5 | 3.0 | 0.1 | 0.75 | 2.3 |

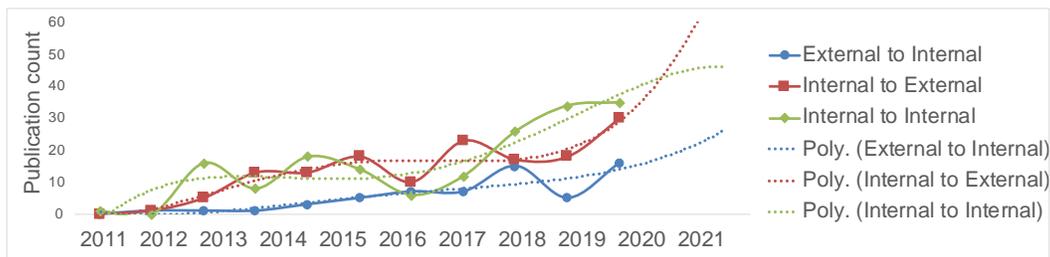

**Figure 17:** Yearly trend of SO data movement between external and internal (SO) data sources.

to the polynomial model, SO content would be used significantly more in the future in different external systems. While this is encouraging and suggests the importance of SO content to the research community, the quality of SO content for external usage is not studied enough. Figure 18 shows the trend of how SO content has been used and studied by the community. Although studies related to SO content analytics and content extraction have been increasing, the number of studies that analyzed SO content quality for external usage was always small.

> **Recommendation:** While researchers can study what other external systems (similar to GitHub) can benefit from SO content, they should also study the quality of SO content before using them for such external systems.

## 6.2. Threats to Validity

It is important to note that some threats can harm the validity of our study.

**Internal validity** is hampered by the authors' bias in searching, selecting, or in the categorization process of the study. We mitigated the possibility of missing papers in two phases. First of all, besides using digital libraries, we also used a search engine (Google Scholar) to collect any residual publications. Secondly, we conducted a paper validation approach described in Section 2.1.4 to validate that our search methodology is robust. Moreover, since there was manual labeling of the facets of the papers by the authors, we conducted an inter-rater reliability test that achieved reasonable agreement as shown in Table 4.

**Construct validity** can be criticized as we have collected SO-based papers to represent Q&A-based crowd knowledge-enhanced research studies. To mitigate this, we analyzed the pervasiveness of SO in software engineering research in Section 3.1 and found that more than 80% of crowd-based SE research is conducted using SO data.

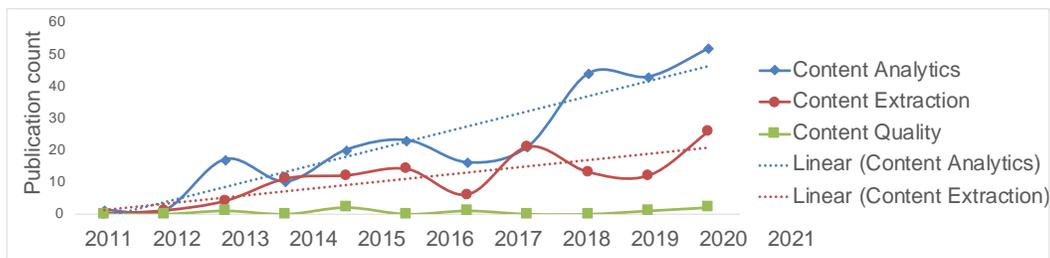

**Figure 18:** Yearly trend of SO content usage

**External validity** refers to the data outside the scope of our study. Crowd-based knowledge generation and research in software engineering could be a pervasive topic. That is why we collected publications ranging from July 2008 to February 2022 (we started analysis of the papers from February 2022) and explicitly used this cut-off period in our search and analysis. Our mapping study cannot validate publications beyond the scope we have defined by the paper collection (in Section 2.1) and exclusion criteria (in Section 2.1.3).

## 7. Related Work

In this section, we first discuss other systematic mapping studies in software engineering. We then focus on the existing systematic mapping studies (or reviews) related to Stack Overflow.

### 7.1. Systematic Mapping Studies in Software Engineering

Systematic mapping studies are popular in software engineering research (e.g., [27, 56, 53, 52, 58, 26, 7, 38, 6, 42, 29, 39, 79, 57, 40], just to name a few). Durelli *et al.* [27] performed a mapping study to understand how machine learning has been used in software testing. They have found that artificial neural networks and decision trees were the most frequently used ML algorithms in software testing. These algorithms were mostly used for test automation, including test Oracle construction, test case generation, and evaluation. Paternoster *et al.* [56] mapped 43 research studies to understand the development practices in different startup companies. Lack of resources and extreme time pressure are the major attributes of these companies. These companies focus less on requirement documentation and often outsource their testing activities. Novais *et al.* [53] mapped research studies related to "software evolution visualization". The authors observed that these visualization studies mainly focused on understanding change history, developing change prediction models, and uncovering the distribution of contributions. There are few mapping studies on different technological concepts. For example, Nuha *et al.* [7] conducted research on microservice architecture, Ramtin *et al.* [38] carried a mapping study on DevOps, Khan *et al.* [40] explored secure software engineering practices, Akoka *et al.* [6] investigated primary research studies on big data. Few mapping studies have been conducted on different life cycle stages of software engineering. Curcio *et al.* [26] conducted a study on requirements engineering in agile development practices. Misaek *et al.* [39] performed a mapping study on non-functional requirements of mobile applications whereas Fernandez *et al.* [29] analyzed usability evaluation methods in web applications. Few mapping studies focused on startup companies as well as organizational maturity models [79, 57].

Cornelissen *et al.* [25] has performed a systematic survey about "program comprehension through dynamic analysis". To systematically label each of the surveyed papers, the authors first developed an attribution framework. Each of the papers was labeled with four different facets (themes): activity, target, method, and evaluation. These facets were then characterized by different attributes. For example, in the activity facet, a paper was labeled as one of the seven attributes: survey, design, views, features, trace analysis, behavior, and general. Encouraged by this systematic approach, we have also developed an *attribution framework* to systematically label all of our 384 SO-based research papers.

### 7.2. Studies Related to StackOverflow

Ahmad *et al.* [5] surveyed 166 research papers to understand the contribution of StackOverflow in software engineering research. The authors, however, focused only on the software development life cycle (SDLC) and found that SO-based research mostly focuses on quantitative analysis. Also, most of the SO papers were limited to a small number of development tasks only. We, however, observed that the contribution of SO in SE research is quite diverse, probably due to the significantly large number of research papers (384 instead of 166) that we include in this paper. In general, our study is significantly different in the following aspects. 1) The study of Ahmad *et al.* was not systematic—they did not follow any systematic approach as was suggested by Kitchenham [41], for example. They rather followed an *ad hoc* paper selection procedure. Also, it was a preliminary study without a set of objectives. For example, no research questions were formed to guide the study. These limitations were mentioned in Section 3.1 of their published version. 2) Unlike their review study, our study is a systematic mapping study, and we have shared our database publicly that categorizes each of the 384 research papers into different facets and attributes. Our shared database can be used by the research community for guiding future research. 3) In the study of Ahmad *et al.*, no research paper was collected after 2016. In this paper, we show that most of the SO-based research papers (69%) were published after 2016. Without this significant number of papers, our understanding of SO's contribution to SE research would be very limited.

Unlike the study of Ahmad *et al.*, a more systematic mapping study of SO-based research was performed by Meldrum *et al.* [49]. The authors found that SO-based studies lack validation, and they are mostly about solution proposals or evaluation. Similar to the study of Ahmad *et al.*, this study also does not contain any research papers after 2016. This is also a very preliminary study (a six-page paper) and does not provide a full picture of how SO has impacted different areas of SE research.

## 8. Conclusion

Among the crowd-sourced community question-answers sites, Stack Overflow stands out in terms of its contribution to software engineering research. Hence, in this mapping study, we explored how the SE research community is utilizing SO in different dimensions of research areas. We have analyzed 384 primary studies published in reputed SE journals and conferences from July 2008 to February 2022 by following rigorous reproducible search methods. We categorized all SO-based studies within ten research facets such as SE domain, SDLC, Research Type, Research

Methodology, and Research Objective. Moreover, we identified the research trend and impact analysis of 18 SE domains that primarily benefited from the SO data.

In general, our shared database can be used by any researcher to find all the Stack Overflow-based software engineering papers within a given area or topic. We also provided specific research areas that can be more useful to the research community for achieving higher impact. Future studies can follow our methodology to produce indexed databases of software engineering research for different sites such as GitHub that share vast information from open-source software projects.

## 9. Acknowledgement

This work was partially supported by the NSERC Discovery Grant (RGPIN/04552-2020) and the NSERC and Alberta Innovates Alliance Grant (ALLRP/568643-2021).